# TECHNOSIGNATURES OF SELF-REPLICATING PROBES IN THE SOLAR SYSTEM

**ALEX ELLERY**
Centre for Self-Replication Research (CESER), Department of Mechanical & Aerospace Engineering, Carleton University, 1125 Colonel By Drive, Ottawa, ON. K1S 5B6. Canada
(aellery@mae.carleton.ca)

## Abstract

We explore a much-neglected area of SETI - solar system technosignatures. As our cursory solar system exploration consolidates into commercial industrialisation, it is crucial that we determine what to look for and where. We first consider the rationale for interstellar self-replicating probes and their implications for the Fermi paradox. Whether for defensive or exploratory reasons, self-replicating probes are a rational strategy for Galactic investigation. We determine that self-replicating probes will systematically explore the Galaxy by tracking resources of sufficient metallicity. We focus on the resource requirements of a self-replicating interstellar probe that may have visited our solar system. After considering asteroid resources, we suggest that evidence of asteroidal processing will be difficult to discern from natural processes given the constraints imposed by self-replication. We further determine that the Moon is an ideal base of manufacturing operations. We suggest that nuclear reactors – such as the Magnox reactor model - can feasibly be constructed from lunar resources which will have left isotopic ratio signatures of $^{232}Th/^{144}Nd$ and/or $^{232}Th/^{137}Ba$. We further suggest that in anticipatory economic trade for resources, a self-replicating probe may have left artefacts buried with asteroidal resources on the Moon. Such gifts would be detectable and accessible only once a threshold of technological sophistication has been achieved. An obvious gift in trade for the resources utilised would be a universal constructor.

## Introduction

In SETI, the traditional search for narrowband radio signatures from extraterrestrial intelligence (ETI) has broadened into a more general search for technosignatures [1,2]. This recent shift has had a long gestation period as there have been numerous proposals for broadening observation modes, for example [3]. Technosignatures make fewer assumptions about ETI motives than the detection of radio signals and potentially permit Kardashev classification [4] based on technosignature where Kardashev index $K=\frac{1}{10}log_{10}W-0.6$ where W=power output of a civilisation=20 x $10^{12}$ W for our current civilisation. For example, a Type II Kardashev civilisation might possess stellar processing capabilities. Blue straggler stars are anomalous phenomena that may be the result of deliberate intervention to prolong their stellar lifetimes, perhaps by an order of magnitude [5]. Such intervention may be achieved using a laser pump to heat the stellar core to drive the circulation of hydrogen in the stellar envelope to its core to stimulate continued nuclear fusion. It would require increasing the stellar hydrogen-burning fraction from 10% to 60%. Blue stragglers have higher luminosities than is characteristic of main sequence stars. However, a natural explanation for blue stragglers is that they result from binary stellar mergers that stimulate excess hydrogen burning. They are found primarily in old metal-poor globular clusters.

---

1 Dorminey B (2008) "Search for astroengineers" *Physics World* **21** (4), 32-35
2 Davis P (2010) "Eerie silence" *Physics World* **23** (3), 28-33
3 Ellery A, Tough A, Darling D (2003) "SETI - a scientific critique and a proposal for further observational modes" *J British Interplanetary Society* **56** (7/8), 262-287
4 Kardashev N (1964) "Transmission of information by extraterrestrial civilisations" *Soviet Astronomy* **8** (2), 217-221
5 Beech M (1990) "Blue stragglers as indicators of extraterrestrial civilisations?" *Earth Moon & Planets* **49**, 177-186

Compared with biosignatures, technosignatures can spread from their biospheric origin across stellar distances and exist independently of their biospheric origins such that $L_{tech}>L_{bio}$ where L=lifetime of biosignature or technosignature [6]. The most obvious vehicle for spreading of such technosignatures is through self-replicating (von Neumann) probes [7, 8]. We briefly review a restricted sample of technosignatures but a broader review is given in [9]. We focus here on megastructures but mention in passing that there are also exoplanet atmospheric technosignatures of infrared absorption lines indicative of industrial pollutants such as $NO_2$, CFCs, etc [10]. The Dyson sphere – a distributed shell of artefacts such as solar power satellites or O'Neill colonies forming a sphere around a star at a distance within the habitable zone to capture the star's energy efficiently - is the archetypal technosignature of a KII civilisation (implying mining and dismantling of asteroids, moons and planets as raw material for constructing the shell) which is detectable as a smooth infrared emission spectrum of its waste heat at 200-400 K peaking at ~10-20 μm [11,12,13]. However, the Dyson sphere technosignature resembles the infrared signature of circumstellar dust enshrouding young stars and red giants. Several searches have been undertaken but no Dyson sphere signatures have been found by IRAS [14,15] or ISO [16]. A KIII civilisation could be characterised by Dyson spheres around every star of a galaxy yielding a galactic infrared signature coincident with optical dimming. Similarly, no such galactic signature has been observed. Planetary-scale artificial structures may be detectable as transit light curves with non-disk features in front of the host star [17] such as a partially-constructed Dyson sphere. The anomalous light curve of Tabby's star, an F-type star 1470 ly away, which has dimmed erratically by as much as 20% over last few years has been occurring for a century. The cause has not yet been determined. Detection of both technosignatures – infrared excess and a non-spherical transiting light curve – would be a strategy that minimises uncertainty in the identification of Dyson spheres [18]. Another technosignature might be artificial structures that may exhibit unnatural orbits, e.g. a distributed solar shield of discrete satellites in

[6] Wright J, Haqq-Misra J, Frank A, Kopparapu R, Lingham M, Sheikh S (2022) "Case for technosignatures: why they may be abundant, long-lived, highly-detectable and unambiguous" *Astrophysical J Letters* **927** (2), L30

[7] Freitas R (1980) "Self-reproducing interstellar probe" *J British Interplanetary Society* **33**, 251-264

[8] Ellery A (2022) "Self-replicating probes are imminent – implications for SETI" *Int J Astrobiology* **21** (4), 212-242

[9] Cariggan R (2010) "Starry messages: searching for signatures of interstellar archaeology" *arXiv:1001.5455v1 [astro-pg.GA]*

[10] Haqq-Misra J, Schwieterman E, Socas-Navarro H, Kopparapu R, Angerhausen D, Beatty T, Berdyugina S, Felton R, Sharma S, de la Torre G, Apai D, TechnoClimes 2020 (2022) "Searching for technosignatures in exoplanetary systems with current and future missions" *Acta Astronautica* **198**, 194-207

[11] Dyson F (1960) "Search for artificial stellar sources of infrared radiation" *Science* **131** (Jun), 1667-1668

[12] Dyson F (1966) "Search for extraterrestrial technology" in *Perspectives in Modern Physics: Essays in Honour of Hans A Bethe on the Occasion of his 60$^{th}$ Birthday*, 641-655

[13] Wright J (2020) "Dyson spheres" *Serbian Astronomical J* **200**, 1-18

[14] Carrigan R (2008) "IRAS-based whole-sky upper limit on Dyson spheres" *Astrophysics J* **698**, 2075-2086

[15] Timofeev Y, Kardashev N, Promyslov V (2000) "Search of the IRAS database for evidence of Dyson spheres" *Acta Astronautica* **46** (10-12), 655-659

[16] Tilgoner C, Heinrichsen I (1998) "Program to search for Dyson spheres with the Infrared Space Observatory" *Acta Astronautica* **47** (10-12), 607-612

[17] Wright J, Cartier K, Zhao M, Jontof-Hutter D, Ford E (2016) "G search for extraterrestrial civilisations with large energy supplies IV. The signatures and information content of transiting megastructures" *Astrophysical J* **816** (1), 17

[18] Teodorani M (2014) "Strategic viewfinder for SETI research" *Acta Astronautica* **105**, 512-516

formation balancing its L1 libration point against light pressure [19]. Motivated to prevent a runaway greenhouse generated as its primary star evolves with increasing luminosity and/or due to $CO_2$ accumulation in its exo-atmosphere, a low-mass starshade of planetary dimensions may be detectable through repeated transits characterised by a maximum blip in the light curve at mid-transit flanked by maximum occultations either side [20]. However, starshades will be detectable only at Earth-sized planets around G stars as a balance between diminutive starshades around late-type stars against the overwhelming luminosity of early-type stars. A near global array of Si photovoltaic panels covering a substantial fraction of the land surface of a terrestrial-type planet may be detectable only with 100s hours integration times to detect its UV/Vis absorption edge (0.34-0.52 μm) [21]. However, a geostationary band of space-based solar power satellites offers superior integrated energy availability while eliminating undesirable thermal effects on climate [22]. Such a band could grow with technological energy demand to form a planetary shell – like a Dyson sphere, it would comprise a constellation of discrete satellites in formation around the planet [23]. This would fit the notion of fleets of geostationary satellites in the Clarke exobelt detectable from a planetary transit light curve ~$10^{-4}L_{*}$ over the primary star exhibiting a steep drop at the edge of the exobelt at a radius of $r=\sqrt[3]{\frac{Gm_pP^2}{4\pi^2}}$ where $m_p$=planet mass, P=period of planet's daily spin [24]. However, the dominance of photovoltaic energy conversion in space is an accident of legacy rather than of efficiency for which there are alternative options such as solar concentrator-thermionic conversion [25,26]. A variation on a planetary shell is the conversion of airless planets into habitable worlds by enclosing them in rigid shells to contain atmospheres that balance gravitational compression against atmospheric tension [27]. Supramundane planets are planetary enclosures around gas giants supported by orbital rings and compression struts offering new landscaped surfaces at a distance where g retains its terrestrial value covered by an atmosphere contained by ~200 km high retaining walls [28]. They would be constructed from outer solar system moons and be differentiable from natural gas giants by their non-hydrogen atmospheric signature.

---

[19] Ellery A (2016) "Low-cost space-based geoengineering: an assessment based on self-replicating manufacturing of in-situ resources on the Moon" *Int J Environmental, Chemical, Ecological, Geological and Geophysical Engineering* **10** (2), S. 278–285

[20] Gaidos E (2017) "Transit detection of a starshade at the inner lagrange point of an exoplanet" *Monthly Notices of Royal Astronomical Society* **469**, 4455-4464

[21] Kopparapu R, Kofman V, Haqq-Misra J, Kopparapu V, Lingham M (2024) "Detectability of solar panels as a technosignature" *Astrophysical J* **967** (2), 119

[22] Ellery A (2016) "Solar power satellites for clean energy enabled through disruptive technologies" *Proc 23rd World Energy Congress (Award Winning Papers)*, Istanbul, Turkey, 133-147

[23] Ellery A (2022) "Solar power satellites – implications of rotary joints" *Proc 73rd Int Astronautics Congress*, Paris, IAC-22.C3.1.x67494

[24] Socas-Navarro H (2018) "Possible photometric signatures of moderately advanced civilisations: the Clarke exobelt" *Astrophysical J* **855**, 110

[25] Ellery A (2016) "Solar power satellites for clean energy enabled through disruptive technologies" *Proc 23rd World Energy Congress (Award Winning Papers)*, Istanbul, Turkey, 133-147

[26] Ellery A (2021) "Generating and storing power on the Moon using in-situ resources" *Proc IMechE J Aerospace Engineering* **236** (6), 1045-1063

[27] Roy K, Kennedy R, Fields D (2013) "Shell worlds" *Acta Astronautica* **82**, 238-245

[28] Birch P (1991) "Supramundane planets" *J British Interplanetary Society* **44**, 169-182

The technosignatures currently proposed are almost entirely astronomical targets, especially stellar and extrasolar planet targets [29]. We suggest that there are potential solar system targets for technosignatures. We adopt the self-replicating probe concept to illustrate that our Moon may be an appropriate location as well as asteroids in the search for technosignatures. There are several principles of forensic science that we may apply to technosignatures [30]: (i) likelihood of an extraterrestrial event is defined as the probability of the event compared with all possibilities (Bayesian approach); (ii) law of progressive change that evidence degrades over time; (iii) principle of exchange that when an extraterrestrial object comes into contact with a terrestrial object, there will be a mutual exchange of material between them; (iv) principle of linkage that physical evidence can be linked to an extraterrestrial cause.

Self-Replicating Probe - Pourquoi et Comment

It has been suggested that there has been a constant increase in ETI civilisations from a galactic age of ~5 until ~10 By (present day) thence levelling out giving a median age of ETI civilisation ~1.7 By [31]. This is constrained by the prevalence of galaxy-sterilising gamma ray bursts during early galactic history up to ~5 By ago. With a median age of 1.7 By for ETI, they, or evidence of their presence, should be here, yet there is a Great Silence [32]. The most obvious motivation to indulge in interstellar exploration for an ancient species lies in its continued survival beyond the main sequence phase of its star. It has been suggested that to counteract the Sun's brightening, Earth's orbit may be migrated outwards using a series of gravity assists of a $10^{19}$ kg main belt asteroid or Kuiper belt object to transfer orbital energy from Jupiter to Earth [33]. This requires ~$10^6$ cycling passes between Jupiter and Earth once every 6000 y to reach 1.5 AU. Alternatively, a large reflective sail (of asteroid-sourced nickel-iron) can use solar radiation pressure to generate thrust and act as a gravity tractor [34]. However, these approaches will not be effective once the Sun shrinks into a white dwarf star. Eventually, interstellar migration will be essential for survival.

The Dark Forest hypothesis attributes the Great Silence to a fearful strategy of self-survival against extermination by other more advanced civilisation [35]. Multiple ETIs will explore quietly – self-replicating probes being the most efficient means to do so - until they are forced to compete for physical resources inevitably leading to conflict, the most technologically superior becoming a singleton. The First Mover to Galactic colonisation will have the advantage of first access of occupation without competition. Simple application of the Prisoner's dilemma to Contact in ignorance of the behaviour of other civilisations (i.e. in the absence of iteration [36]) imposes a Berserker strategy of the destruction of the Other. Given this defensive strategy, comprehensive

[29] Haqq-Misra J, Schwieterman E, Socas-Navarro H, Kopparapu R, Angerhausen D, Beatty T, Berdyugina S, Felton R, Sharma S, de la Torre G, Apai D, TechnoClimes 2020 (2022) "Searching for technosignatures in exoplanetary systems with current and future missions" *Acta Astronautica* **198**, 194-207

[30] Nickel J, Fischer J (1999) *Crime Science: Methods of Forensic Detection*, University Press of Kentucky, USA

[31] Norris R (2000) "How old is ET?" in *When SETI Succeeds: The Impact of High-Information Contact* (ed. Tough A), Foundation for the Future, Washington DC

[32] Brin D (1983) "Great Silence – the controversy concerning extraterrestrial intelligent life" *Quarterly J Royal Astronomical Society* **24** (3), 283-309

[33] Korycansky D, Laughlin G, Adams F (2001) "Astronomical engineering: a strategy for modifying planetary orbits" *Astrophysics & Space Science* **275**, 349-366

[34] McInnes C (2002) "Astronomical engineering revisited: planetary orbit modification using solar radiation pressure" *Astrophysics & Space Science* **282**, 765-772

[35] Yu C (2015) "Dark forest rule: one solution to the Fermi paradox" *J British Interplanetary Society* **68**, 142-144

[36] Axelrod R (1984) *Evolution of Cooperation*, Basic Books, USA

exploration of the interstellar environment is essential to gain military intelligence for situational awareness [37]. Situational awareness is fundamental to any generalised cognitive perception-action cycle in which perception of the situation (performance) informs adaptive action (competence) on the environment in relation to risk, be it terrestrial or extraterrestrial. Taking a more optimistic stance, according to Dick's "Intelligence Principle", a technical culture will undergo cultural evolution to grow its knowledge [38]. Cultural evolution in long-lived species may evolve from biological intelligence to postbiological intelligence implemented as machine intelligence [39]. Galactic simulations of exploration probes suggest that the number of probes currently permeating the Galaxy is capped at ~1000 assuming a 1 My probe lifetime and that as the latter increases to 100 My so the number of extant probes drops to ~10 [40]. This neglects self-replicating probes that can permeate the Galaxy rapidly and completely. If there are multiple extraterrestrial civilisations, no matter their relative temporal emergence, they could each send out self-replicating probes so the Galaxy should be saturated with such probes [41]. Self-replicating probes are an obvious vehicle for cultural evolution and growth because they maximise knowledge acquisition to the sending civilisation by exploring the Galaxy systematically, exhaustively and completely. If every human life is a good, then expansion of the "human" species, in whatever form that may be, increases good. If we assume that our own Virgo galactic supercluster of $\sim 10^{13}$ stars can support $\sim 10^{23}$ humans ($\sim 10^{10}$ humans/star) then delays in space colonisation imposes an opportunity cost of the loss of $\sim 10^{14}$ potential human lives per second [42]. Some estimates of loss of future human life are even higher at $>5\times10^{46}$ if human galactic colonisation is delayed by 100 y by decisions on space exploration made today [43]. The arguments in favour of interstellar expansion are thus several and profound. The self-replicating interstellar probe is the means through which to achieve galactic colonisation and beyond. The technological feasibility of self-replicating probes has been discussed extensively in [44]. The duration between the establishment of radio communications and the development of interstellar flight (contact era) has been suggested as a few hundred to a few thousand years so that the expanding radiosphere is a more likely technosignature than physical probes [45]. This does not account for exponential technological growth particularly with the advent of self-replication technology emerging on Earth [46] within a century of radio technology as a near-term capability for humanity [47, 48] – indeed, self-replication technology may be developed prior to interstellar

[37] Smith K, Hancock P (1995) "Situation awareness is adaptive, externally directed consciousness" *Human Factors* **37** (1), 137-148

[38] Dick S (2008) "Postbiological universe" *Acta Astronautica* **62**, 499-504

[39] Dick S (2003) "Cultural evolution, the postbiological universe and SETI" *Int J Astrobiology* **2** (1), 65-74

[40] Cotta C, Morales A (2009) "Computational analysis of galactic exploration with space probes: implications for the Fermi paradox" *arXiv: 0907.0345v1 [physics.pop-ph]*

[41] Wiley K (2011) "Fermi paradox, self-replicating probes and the interstellar transportation bandwidth" *arXiv: 1111.6131v1 [physics.pop-ph]*

[42] Bostrom N (2003) "Astronomical waste: the opportunity cost of delayed technological development" *Utilitas* **15** (3), 308-314

[43] Cirkovic M (2002) "Cosmological forecast and its practical significance" *J Evolution & Technology* **12** (1)

[44] Ellery A (2022) "Self-replicating probes are imminent – implications for SETI" *Int J Astrobiology* **21** (4), 212-242

[45] Wandel A (2022) "Fermi paradox revisited: technosignatures and the contact era" *Astronomical J* **941**, 184

[46] Ellery A (2016) "Are self-replicating machines feasible?" *AIAA J Spacecraft & Rockets* **53** (2), 317-327

[47] Ellery A (2018) "Life without organic molecules – exploring the boundaries of life" *The Biochemist* **40** (6), 14-17

[48] Ellery A (2020) "How to build a biological machine using engineering materials and methods" *Biomimetics J* **5** (3), 35

propulsion technology which itself is expected to be a practical proposition within a half-century [49] for our K0.73 civilisation.

The Great Filter [50] was premised on anthropic "hard steps" of biological evolution [51] which has been subject to criticism [52] but its application to future technological hurdles is not subject to such criticism. Artificial intelligence (AI) evolving into advanced superintelligence (ASI) prior to the establishment of multiplanetary settlements [53] has been suggested as a Great Filter [54]. Sophisticated terrestrial AI systems based on deep neural networks have emerged rapidly and have become ever more ubiquitous. However, the current AI paradigm based on transformer networks implementing large language models [55] are suited to the human social world but not the larger physical world as they are not embedded in the sense of sensorimotor interaction [56], rendering it an unlikely platform for ASI. For applications in starships operating in an unknown physical world [57], symbolic-neural AI is more suitable for embedding in the physical world [58,59] through symbol grounding [60] and this may potentially offer a route to ASI. The evolution of ASI that interacts with the physical world could threaten humanity as a competitor for physical resources. This stresses the urgency for human development of space technology within a short temporal window [61] – self-replication technology provides such exponential leverage for space industrialisation [62]. There is, then, a nuance in which a sufficiently capable AI is necessary for a self-replicating probe to function but which must be constrained in its broader capabilities to impose harm – an ethical AI [63]. There are arguments that self-replication technology itself does not constitute a Great Filter [64] and, if the Great Filter is subsequent to the development of these technologies, the self-replicating probe renders it irrelevant to the Fermi paradox. Self-replicating probes must implement AI to function

---

[49] Lubin P (2016) "Roadmap to interstellar flight" *J British Interplanetary Society* **69**, 40-72
[50] Hanson R (1998) "The Great Filter – are we almost past it?" *https://mason.gmu.edu/~rhanson/greatfilter.html*
[51] Barrow J, Tipler F (1988) *Anthropic Cosmological Principle*, Oxford University Press, UK
[52] Mills D, Macalady J, Frank A, Wright J (2025) "Reassessment of the hard steps model for the evolution of intelligent life" *Science Advances* **11** (Feb), eads5698
[53] Garrett M (2024) "Is artificial intelligence the great filter that makes advanced technical civilisations rare in the universe?" *Acta Astronautica* **219**, 731-735
[54] Hanson R (1998) "The Great Filter – are we almost past it?" *https://mason.gmu.edu/~rhanson/greatfilter.html*
[55] Hutson M (2021) "Language machines" *Nature* **591** (Mar), 22-25
[56] Barsalou L (2008) "Grounded cognition" *Annual Reviews Psychology* **59**, 617-645
[57] Ellery A (2025) "State of hybrid artificial intelligence for interstellar missions" *Progress in Aerospace Sciences* **156**, 101100
[58] Ellery A (2015) "Artificial intelligence through symbolic connectionism – a biomimetic rapprochement" in *Biomimetic Technologies: Principles & Applications* (ed. Ngo D), Elsevier Publishing
[59] Ellery A (2010) "Selective snapshot of state-of-the-art artificial intelligence and robotics with reference to the Icarus starship" *J British Interplanetary Society* **62**, 427-439
[60] Harnad S (1990) "Symbol grounding problem" *Physica D: Nonlinear Phenomena* **42** (1-3), 335-346
[61] Garrett M (2024) "Is artificial intelligence the great filter that makes advanced technical civilisations rare in the universe?" *Acta Astronautica* **219**, 731-735
[62] Ellery A (2024) "Self-replication technology for ubiquitous space exploration for all" *Int Astronautical Congress*, Milan, IAC-24-D4.1.9x80842
[63] Ellery A (2024) "Self-replicating probes are a reliable strategy for ETI" *Int Astronautical Congress*, Milan, IAC-24-A4.1.14x80843
[64] Ellery A (2024) "Self-replicating probes are a reliable strategy for ETI" *Int Astronautical Congress*, Milan, IAC-24-A4.1.14x80843

effectively and Contact with ETI within our solar system, if it does indeed occur, will most likely be with machine, rather than biological, intelligence [65].

Self-replicating probe expansion into the Galaxy is robust to the interstellar propulsion method adopted with only a narrow set of orders of magnitude ~$10^2$ variance in Galactic colonisation time [66]. There are unbound stellar pairs within 100 pc of the Sun that undergo close encounters within ~$10^4$ AU of each other providing short interstellar migration routes for technological civilisations [67]. They require only modest interstellar transport requirements but more demanding interstellar travel requires significant interstellar propulsion technology. Zubrin has characterised emitted radiation technosignatures of four modes of interstellar transport [68]: (i) antimatter rocketry; (ii) fusion rocketry; (iii) fission rocketry; (iv) magnetic sails. Gamma ray emissions from antimatter, fusion and fission are at different energies but would be undetectable at interstellar distances. Visible light emitted from antimatter drives are detectable telescopically at ~100 ly in the line of sight of the exhaust. Bremsstrahlung X-ray emissions from fusion engines are detectable at ~1 ly, less than typical interstellar distances. Cyclotron radio emissions from magnetic braking is detectable at ~$10^3$ ly. Extragalactic fast radio bursts could be interstellar sail microwave drivers at ~1 GHz although they are, less extraordinarily, believed to be generated by magnetars [69]. Of course, these propulsive signals are detectable only during acceleration/deceleration which are of short duration compared with the cruise phase of an interstellar ship.

The single most important characteristic of civilisation is economic growth P(t) implying exponential energy growth $\dot{E}$ [70]:

$$\dot{E} = \lambda \int_0^t P(t).dt$$

where $\lambda = \frac{\dot{E}}{\int_0^t P(t).dt}$ =constant (W/$). The integral of P(t) through time from the start of civilisation constitutes wealth where economic growth rate is given by:

$$r = \frac{P}{\int_0^t P(t).dt}$$

Energy consumption drives economic growth as $f = \frac{P}{\dot{E}}$ so economic growth rate $r = \lambda f = \frac{1}{\dot{E}}\frac{d(\dot{E})}{dt}$. Economics as the rational distribution of resources is a fundamental constraint. The argument that interstellar communication is cheaper than interstellar travel [71] does not apply in the context of a self-replicating probe because the initial capital cost of the first self-replicating probe is amortised over subsequent generations [72] – these subsequent generations are economically free to the

---

65 Rees M (2021) "SETI: why extraterrestrial intelligence is more like to be artificial rather than biological" *The Conversation*, 18 April 2021

66 Matloff G (2022) "von Neumann probes: rationale, propulsion, interstellar transfer timing" *Int J Astrobiology* **21** (4), 205-211

67 Hansen B (2022) "Unbound close stellar encounters in the solar neighbourhood" *Astronomical J* **163**, 44

68 Zubrin R (1996) "Detection of extraterrestrial civilisations via the spectral signature of advanced interstellar spacecraft" *J British Interplanetary Society* **49**, 297-302

69 Lingham M, Loeb A (2017) "Fast radio bursts from extragalactic light sails" *Astrophysical J Letters* **837**, L23

70 Garrett T (2011) "How persistent is civilization growth?" *arXiv:1101.5635v1 [physics.soc-ph] 28 Jan 2011*

71 Scheffer L (1994) "Machine intelligence, the cost of interstellar travel and Fermi's paradox" *QJ Royal Astronomical Society* **35**, 157-175

72 Ellery A (2017) "Space exploration through self-replication technology compensates for discounting in NPV cost-benefit analysis – a business case?" *New Space J* **5** (3), 141-154 sending civilisation [73]. One assumption is that exponential civilisation growth is not sustainable on a galactic scale, so there has been insufficient time for ETI to have expanded through the Galaxy [74]. As this is irrelevant to self-replicating probes, a corollary of this argument as a solution to the Fermi paradox is that self-replicating probes are impossible. As long as the replication rate exceeds the failure rate, expansion of self-replicating probes continues [75]. Swarms of self-replicating microprobes ~0.1 mm in size may be detectable as non-thermal infrared signatures whilst passing through and mining HII regions assuming KII- and KIII-scale energy emissions [76,77]. Billions of nano- or pico-gram-scale interstellar probes propelled by graphene light sails ~$10^{-8}$ $m^2$ through interstellar space could expand their sail diameter to decelerate and land on habitable planets or moons with liquid water [78]. They then self-replicate to construct planetary scale luciferase-based bioluminescent arrays whose emissions may be pulse-controlled to encode messages observable by telescope. Perhaps, synthetic biology may permit the exploitation of the cyanobacterium *Synechocystis* whose cells act as microlenses [79] to amplify such signals. However, we consider macroscale self-replicating probes to be a more feasible technology within immediate human capacities [80]. It has been argued that, given the conceptual similarity between von Neumann's universal constructor and Turing's universal computer, the self-replicating machine emerges as an attempt to solve the halting problem for universal construction [81]. The self-replication process therefore never halts because it is undecidable. A corollary of this argument is that self-replicating probes would voraciously consume the Galaxy within 150 generations over ~My into ~$10^{47}$ self-replicating probes that would then cannibalise each other [82]. Such arguments are unwarranted as biological extinctions clearly demonstrate that self-replication can be terminated. Furthermore, the prevention of mutations that drive evolutionary divergence of self-replicating probes can be implemented through a multi-layered strategy of error detection and correction coding [83] and the implementation of Hayflick limits on generation number [84]. Such genetic integrity also eliminates any cumulative error catastrophe suggested as a reason why self-replicating probes are unfeasible [85]. Similarly, manufacturing integrity of the self-replicating probe over generations may be

---

[73] Tipler F (1980) "Extraterrestrial intelligent beings do not exist" *Quarterly J Royal Astronomical Society* **21**, 267-281

[74] Haqq-Misra J, Baum S (2009) "Sustainability solution to the Fermi paradox" *J British Interplanetary Society* **62**, 47-51

[75] Ashworth S (2022) "Self-replicating interstellar probes and runaway growth reconsidered" *J British Interplanetary Society* **75** (8), 283-292

[76] Osmanov Z (2020) "On interstellar von Neumann micro self-reproducing probes" *Int J Astrobiology* **19** (3), 220-223

[77] Osmanov Z (2020) "On a spectral pattern of von Neumann probes" *J British Interplanetary Society* **73**, 254-257

[78] Church G (2022) "Picogram-scale interstellar probes vis bioinspired engineering" *Astrobiology* **22** (12), 1452-1458

[79] Nilsson D-E, Colley N (2016) "Comparative vision: can bacteria really see?" *Current Biology* **26**, R355-R371

[80] Ellery A (2022) "Self-replicating probes are imminent – implications for SETI" *Int J Astrobiology* **21** (4), 212-242

[81] Sayama H (2008) "Construction theory, self-replication and the halting problem" *Complexity* **13** (5), 16-22

[82] Sagan S, Newman W (1983) "Solipsist approach to extraterrestrial intelligence" *Quarterly J Royal Astronomical Society* **24**, 113-121

[83] Ellery A, Eiben G (2019) "To evolve or not to evolve: that is the question" *Proc Artificial Life Conf*, 357-364

[84] Ellery A (2022) "Curbing the fruitfulness of self-replicating machines" *Int J Astrobiology* **21** (4), 243-259

[85] Kowald A (2015) "Why is there no von Neumann probe on Ceres? Error catastrophe can explain the Fermi-Hart paradox" *J British Interplanetary Society* **68**, 383-388

maintained through precision measurement and calibration [86]. Genetic integrity would be essential to prevent an evolutionary arms race between co-evolving predator and prey probes [87]. In any case, even if self-replicating probes evolved into predator-prey populations, Lotka-Volterra model dynamics generates oscillatory populations (limit cycle with period $T = 2\pi\sqrt{\tau_{prey}\tau_{pred}}$ where τ=predator/prey characteristic replication rates) rather than extinction of self-replicator populations [88,89]. The Fermi paradox then remains extant [90].

We may examine the spread of self-replicating probes through the Galaxy. Density-dependent interstellar migration may be modelled as r- and K-strategies such that Malthusian fitness m is given by $m = \frac{dN}{dt} = r - \left(\frac{r}{K}\right)N$ where N=population, r=population growth rate, K=environment carrying capacity [91]. Given that K is dependent on local resource availability which is far greater than that required for self-replication, the self-replicating probe would indulge in an r-strategy of exploration rather than a K-strategy since carrying-capacity K of the extrasolar system is irrelevant to the self-replication rate r of such probes. The K-strategy would be applicable only to much slower worldships [92] – such as mobile O'Neill colonies [93] - in which human (or alien) populations are transported to inhabit new worlds [94] which eventually impose resource saturation due to population pressure. Limited migration rates between interstellar colonies could impose cultural isolation between such colonies [95] but this has no causal effect on population pressure to emigrate to new worlds. Self-replicating probes would be essential to precede worldships for prior surveying to select appropriate worldship destinations with appropriate resources to reduce risk to its passengers. Self-replication technology will also be essential for the construction of worldships to minimise their costs, as a fabrication payload for self-repair of the worldship during transit, and for the construction of O'Neill colonies at the target destination to accommodate a population and its subsequent growth [96]. Self-replicating probes are independent of worldship colonisation so directed exploration (r-strategy) [97] is appropriate rather than diffusional or percolation exploration motivated by population pressure

---

[86] Winchester S (2019) *The Perfectionists: How Precision Engineers Created the Modern World*, Harper Perennial Publishers

[87] Nolfi S, Floreano D (1998) "Co-evolving predator and prey robots: do arms races arise in artificial evolution" *Artificial Life* **4** (4), 311-335

[88] Allen P (1975) "Darwinian evolution and a predator-prey ecology" *Bulletin of Mathematical Biology* **37**, 389-405

[89] Allen P (1976) "Evolution, population dynamics and stability" *Proc National Academy Sciences* **73** (3), 665-668

[90] Chen Y, Ni J, Ong C (2022) "Lotka-Volterra models for extraterrestrial self-replicating probes" *European Physical J Plus* **137**, 1109

[91] Boyce M (1984) "Restitution of r- and K-selection as a model of density-dependent natural selection" *Annual Review Ecological Systems* **15**, 427-447

[92] Martin A (1984) "World ships – concept, cause, cost, construction and colonisation" *J British Interplanetary Society* **37**, 243-253

[93] O'Neill G (1977) *The High Frontier: Human Colonies in Space*, William Morrow & Co, USA

[94] Hein A, Pak M, Putz D, Buhler C, Reiss P (2012) "World ships – architectures and feasibility" *J British Interplanetary Society* **65**, 119-133

[95] Wiley K (2011) "Fermi paradox, self-replicating probes and the interstellar transportation bandwidth" *arXiv:1111.6131v1 [physics.pop-ph] 26 Nov 2011*

[96] Lamontagne M (2021) "Worldship and self-replicating systems" *Principium* **32** (Feb), 29-43

[97] Jones E (1976) "Colonisation of the Galaxy" *Icarus* **28** (3), 421-422

(K-strategy) [98,99]. Percolation theory's main parameters are the fixed degree of connectivity, fixed probability of propagation to neighbouring star systems and lifetime of a colony [100]. This restricts colonisation efficiency. More sophisticated settlement models including diffusive stellar motions and finite settlement lifetimes suggest that there will be clusters of settled worlds interspersed by unvisited regions and the Earth could reside within an unvisited region surrounded by interstellar civilisations [101]. Settlements spread across the Galaxy rapidly but limited settlement lifetimes with retarded interstellar probe launches impose these voids. The light cage model assumes that civilisation collapse due to exponential population growth and resource exhaustion within 8000 y cannot be outpaced by a colonisation wavefront even as $v \rightarrow c$ which has a maximum radius of 300 ly [102]. Exponential population growth in humans, although historically accurate, is slowing down and projected to peak at 10 billion around 2080. Indeed, rapid growth may be unsustainable on a galactic scale and that exponentially growing civilisations collapse [103]. Such collapse occurs due to resource exhaustion, environmental degradation and conflict. Slower growing civilisations that husband their resources and protect their environment thrive but expand only slowly. All these models are models of colonisation patterns driven by population pressure. They are not models of exploration which are more aggressive patterns [104] which, for a self-replicating probe, would be an exponential breadth-first search guaranteed to cover all space constrained only by resource availability rather than settleability. Thus, the Fermi paradox stands. The Fermi paradox argument regarding self-replicating probes may be expanded to incorporate other galaxies [105].

Self-Replicating Probe Follows the Crumbs

The Galactic habitable zone (GHZ) is an annular band of ~$1.2 \times 10^{10}$ stars ~7-9 kpc from the Galactic Centre when stellar metallicity $Z > Z_\odot$ was sufficient to form terrestrial planets [106]. It formed ~4-8 By ago to be of sufficient age for biological evolution and sufficient distance from frequent, dense supernovae, active galactic nuclei and gamma ray bursts. As metallicity increased over time, the GHZ increased outward. It is a thin disk ~600 pc thick of habitable stars that excludes the inner bulge, halo and thick disk. Our Sun is located centrally in the GHZ at 8 kpc near the galactic plane. The inclusion of radial gas inflows increases the size of the GHZ to 8-12 kpc [107]. A more detailed

98 Newman W, Sagan C (1981) "Galactic civilisations: population dynamics and interstellar diffusion" *Icarus* **46** (3), 293-327

99 Cartin D (2014) "Quantifying the Fermi Paradox in the local solar neighbourhood" *J British Interplanetary Society* **67**, 119-126

100 Landis G (1998) "Fermi paradox: an approach based on percolation theory" *J British Interplanetary Society* **51**, 163-166

101 Carroll-Nellenback J, Frank A, Wright J, Scharf C (2019) "Fermi paradox and the aurora effect: exo-civilisation settlement, expansion and steady states" *Astronomical J* **158** (3), 117

102 MacInnes C (2002) "Light cage limit to interstellar expansion" *J British Interplanetary Society* **55**, 279-284

103 Haqq-Misra J, Baum S (2009) "Sustainability solution to the Fermi paradox" *J British Interplanetary Society* **62**, 47-51

104 Jones E (1981) "Discrete calculations of interstellar migration and settlement" *Icarus* **46**, 328-336

105 Armstrong S, Sandberg A (2013) "Eternity in six hours: intergalactic spreading of intelligent life and sharpening the Fermi Paradox" *Acta Astronautica* **89**, 1-13

106 Lineweaver C, Fenner Y, Gibson B (2004) "Galactic habitable zone and the age distribution of complex life in the Milky Way" *Science* **303** (Jan), 59-62

107 Spitoni E, Matteucci F, Sozzetti A (2014) "Galactic habitable zone of the Milky Way and M31 from chemical evolution models with gas radial flows" *Monthly Notices Royal Astronomical Society* **440**, 2588-2598 asymmetric 3D map of the GHZ increases the broadness of its range to 2-13 kpc [108]. Some 75% of stars in the GHZ are older than the Sun at ~5 By implying the average age of stars in the GHZ is ~1 By older than the Sun and that ETI is more likely to be ancient, consistent with Norris' estimate [109]. It has been suggested that postbiological machine intelligence will favour colder regions in the outer Galactic disk for efficient computation constrained by metallicity to supply their need for metal resources that diminish in the outer Galactic rim [110] but this "computational" zone is unnecessary as thermal energy can be radiated to the 3K cosmic microwave background of deep space from anywhere. Sufficient metallicity for the emergence of life is different to that required for resource extraction by a self-replicating probe. There is a minimum metallicity for planet formation $[Fe/H]_{crit}\approx$-1.5+log(r/1 AU), i.e. Z$\geq 0.1Z_\odot$ where $Z_\odot$=solar metallicity=0.02 [111]. Supernovae of first-generation population III stars enriched metallicity of the second-generation population II stars to $Z{\sim}10^{-3}Z_\odot$. It reached $Z{\sim}0.1Z_\odot$ after several generations of population II stars within the first ~1 By of Galaxy formation. The frequency of low mass (0.3-0.7$M_\odot$) star formation of low metallicities in the past was diminished compared with that predicted by the Salpeter initial mass function (IMF) given by $p(m_*)dm_* \propto m_*^{-\alpha}$ where $m_*$=stellar mass, α=2.35 [112]. This suggests that the vast majority of stars in the Galaxy's outer thin disk (metallicity distribution peak at ~$Z_\odot$) and inner thick disk (metallicity distribution peak at ~$0.2Z_\odot$) possess planetary bodies that are suitable for raw material exploitation and resource recovery. Galactic colonisation of $\geq 0.1Z_\odot$ stars will thus proceed to completion. Furthermore, we expect asteroidal resources in extrasolar systems to be comparable to our own. Presolar interplanetary dust particles include metal nitrides, carbides, oxides, silicates and FeNi alloys [113]. Such dust particles accrete into planetesimals during planet formation. The detection of silicon monoxide gas and condensed silicate mineral (silica, forsterite and enstatite) crystals in the inner accretion disk around a young protostar suggests the generality of the temperature condensation sequence of the solar nebula [114]. Accretion of extrasolar planetesimals onto several white dwarfs within 200 pc indicate that the dominant constitution – O, Mg, Si and Fe and depletion of volatiles C, S and N - is similar to that of our own asteroids and moons but with higher concentrations of elements heavier than Ca [115]. This suggests that geochemical distribution and evolution is broadly universal and that the Bowen condensation sequence from magma is an appropriate model of predictive extraterrestrial mineralogy (Fig 1).

| Mafic series | Felsic series |
|---|---|
| Olivine | Calcium feldspar |
| \ | (anorthite) |

[108] Forgan D, Dayal P, Cockell C, Libeskind N (2017) "Evaluating galactic habitability using high resolution cosmological simulations of galaxy formation" *Int J Astrobiology* **16** (1), 60-73

[109] Norris R (2000) "How old is ET?" in *When SETI Succeeds: The Impact of High-Information Contact* (ed. Tough A), Foundation for the Future, Washington DC

[110] Cirkovic M, Bradbury R (2006) "Galactic gradients, postbiological evolution and the apparent failure of SETI" *New Astronomy* **11**, 628-639

[111] Johnson J, Li H (2012) "First planets: the critical metallicity for planet formation" *Astrophysical J* **751**, paper no 81

[112] Li J, Liu C, Zhang Z-Y, Tian H, Fu X, Li J, Yan Z-Q (2023) "Stellar initial mass function varies with metallicity and time" *Nature* **613** (Jan), 460-462

[113] McCoy T (2010) "Mineralogical evolution of meteorites" *Elements* **6** (Feb), 19-23

[114] McClure M, van't Hoff M, Francis L, Bergin E, Rocha W, Sturm J, Harsono D, van Dishoeck E, Black J, Noble J, Qasim D, Dartois E (2025) "Refractory solid condensation detected in an embedded protoplanetary disk" *Nature* **643** (Jul), 649-653

[115] Jura M, Xu S, Klein B, Koester D, Zuckerman B (2012) "Two extrasolar asteroids with low volatile-element mass fractions" *Astrophysical J* **750**, article 69

```
Pyroxene                      /
     \                       /
     Amphibole              /
       \                   /
        Biotite mica     Sodium feldspar
         \               (albite)
          \             /
           Potassium feldspar
                  |
           Muscovite mica
                  |
               Quartz
```

Fig 1. Bowen's reaction series

Self-Replicating Probe Mission Profile

We shall assume that the self-replicating probe's priorities will be: (i) targeting ubiquitous asteroids and moons for readily-accessible raw materials required for universal construction; (ii) build surveying probes to comprehensively survey the extrasolar system for resources and life-bearing environments: (iii) select and secure the primary locations for basing self-replication operations: (iv) self-replicate; (v) long-term, detailed exploration of the extrasolar system using surveying probes as sentinels; (vi) specific task instructions such as building O'Neill colonies without encroaching on potentially life-bearing planets, and/or more controversially, seeding life on potentially life-bearing planets (directed panspermia [[116]]). If self-replicating probes have visited Earth or our solar system, there should be evidence of such within our solar system – indeed, the search for solar system technosignatures has been neglected compared with astronomical searches for technosignatures [[117]]. One of the chief problems is that the probability of a recent visit by ETI is vanishingly small so any artefacts of such visitation in the distant past must be extremely stable [[118]]. This excludes the Earth itself (which has been extensively explored and subject to dynamic orogenic, tectonic and erosional processes though there may exist geologically ancient technosignatures [[119]]) but still permits siting on other bodies in the solar system that are tectonically stable, although destruction by impact hazards cannot be excluded over aeons [[120],[121]]. Similarly, for any early indigenous technological species that may have arisen on early Mars when it was wet or Venus prior to its runaway greenhouse, technosignature evidence would be unlikely to survive [[122]].

Targeting Asteroids

[116] Crick F, Orgel L (1973) "Directed panspermia" *Icarus* **19** (3), 341-346

[117] Haqq-Misra J, Schwieterman E, Socas-Navarro H, Kopparapu R, Angerhausen D, Beatty T, Berddyugina S, Felton R, Sharma S, de la Torre G, Apai D, TechnoClimes Workshop participants (2022) "Searching for technosignatures in exoplanetary systems with current and future missions" *Acta Astronautica* **198**, 194-207

[118] Davies P (2012) "Footprints of alien technology" *Acta Astronautica* **73**, 250-257

[119] Schmidt G, Frank A (2019) "Silurian hypothesis: would it be possible to detect an industrial civilisation in the geological record?" *Int J Astrobiology* **18** (2), 142-150

[120] Chapman C, Morrison D (1994) "Impacts on the Earth by asteroids and comets: assessing the hazards" *Nature* **367**, 33-40

[121] Chapman C (2004) "Hazard of near-Earth asteroid impacts on Earth" *Earth & Planetary Science Letters* **222**, 1-15

[122] Wright J (2018) "Prior indigenous technological species" *Int J Astrobiology* **17** (1), 96-100

For a self-replicating probe, the first task requires selection of high-resource targets. Asteroids are expected to have a gamut of resources for feeding the self-replicating machine – from iron-nickel-cobalt and other metals to volatiles such as carbon, nitrogen, sulphur and water. It has been suggested that the spectral technosignatures from deliberate asteroid mining in extrasolar systems will be indistinguishable from infrared/submillimetre spectra of natural debris disks and collisional processes [[123]]. If so, our own asteroid belt and moons may be a more likely search target for success. The asteroid belt should be explored because of its abundant and varied mineral resources while the Kuiper belt and Oort cloud are unlikely candidates as repositories dominated by frozen volatiles (although useful as a hydrogen fuel source) and a lack of solar energy availability – even if nuclear or other energy sources are exploited, solar energy provides an efficient ~90% source of direct high-temperature thermal energy ~1500°C using Fresnel lens concentrators. The high-inclination Centaurs and trans-Neptunian objects which have been nearly polar for ~4.5 By suggests that they may have been captured from the interstellar medium [[124]]. They represent opportunistic targets for interstellar hopping but the same criticism of distance from the Sun applies as resources. It is plausible that a communications relay may be emplaced at a star's gravitational lens focus (for the Sun >550 AU) that can significantly increase the gain ~120 dB of interstellar transmission along the Sun-focus-target vector, perhaps as part of an interstellar network [[125]]. The most significant perturbations that must be countered through active propulsion include inward gravity of the sun and planetary influences ~10 m/s/y necessitating refuelling every ~5000 y. It has been suggested that interstellar comets outnumber solar system comets in the Oort cloud at 2000-100,000 AU representing a vast source of hydrogen, carbon and oxygen [[126]]. However, this is effectively unreachable from the solar focus at >550 AU and the interstellar comets are gravitationally unbound with rapid characteristic velocities of ~30 km/s. The nearest fuel resource would be the Kuiper belt at 30-50 AU requiring a Hohmann transfer Δv of 1.2 km/s with a flight time of 2500 y, i.e. a cycling orbit. This implies a considerable infrastructure. This would be difficult to detect at such a distance from Earth but such a communications relay pre-supposes exploratory probes as transmitters of data.

The main asteroid belt at 2.8 AU (inner belt) and 3.2 AU (outer belt) offer much higher solar availability. Its size-frequency distribution (SFD) has been in a quasi-stable collisional steady state for ~4 By with a collision probability of 2.9 x $10^{-18}$/$km^2$/y at an impact velocity of 5.3 km/s [[127]]. This suggests that interstellar probes lurking in the asteroid belt would have been stable over astronomical periods of time. The asteroid belt is an appropriate location for extraterrestrial asteroid mining facilities [[128]]. We select A-type asteroids as representative of both M-type and S-type in that they comprise a median species with 50% olivine ($Mg_2SiO_4$) and pyroxene (dominantly enstatite $MgSiO_3$ also representative of E-type asteroids) silicate grains embedded in a 50% matrix of Fe-Ni-Co metal alloy with some FeS inclusions. The key to any technosignature resides in the discarded

---

[123] Forgan D, Elvis M (2011) "Extrasolar asteroid mining as forensic evidence for extraterrestrial intelligence" *Int J Astrobiology* **10** (4), 307-313

[124] Namouni F, Morais M (2020) "Interstellar origin for high-inclination Centaurs" *Monthly Notices Royal Astronomical Society* **494**, 2191-2199

[125] Kerby S, Wright J (2021) "Stellar gravitational lens engineering for interstellar communication and artefact SETI" *Astronomical J* **162**, 252

[126] Siraj A, Loeb A (2021) "Interstellar objects outnumber solar system objects in the Oort cloud" *Monthly Notices Royal Astronomical Society* **507**, L16-L18

[127] Bottke W, Broz M, O'Brien D, Bagatin C, Morbidelli A, Marchi S (2015) "Collisional evolution of the main asteroid belt" in *Asteroids IV* (eds. Michel P et al), University of Arizona Press, 701-724

[128] Papagiannis M (1983) "Importance of exploring the asteroid belt" *Acta Astronautica* **10** (10), 709-712

waste products that are of little use and so remain unconsumed. Searching for evidence of processing requires insight into the type of processing that might be performed. The discovery of extensive waste deposits that would be unexpected naturally would invite artificial explanation. The major constraint in closing the self-replicator loop for a self-replicating probe requires [[129]]: (i) minimising waste (sustainability) to minimise energy consumption; (ii) minimising materials and components manufacture to minimise mining and chemical processing; (iii) minimising manufacturing and assembly processes to minimise infrastructure machinery.

The most obviously useful materials are the M-type metal alloys. The carbonyl process can convert metal (Fe, Ni or Co) into its carbonyl form under different but modest conditions thereby separating them, all involving a FeS catalyst [[130]]:

$Fe(CO)_5 \leftrightarrow 5CO + Fe$ (175°C/100 bar)

$Ni(CO)_4 \leftrightarrow 4CO + Ni$ (55°C/1 bar)

$Co_2(CO)_8 \leftrightarrow 8CO + 2Co$ (150°C/35 bar)

Processing of M-type asteroids for their iron-nickel-cobalt metal alloys would involve the reversible carbonyl process to separate the individual metals. In their gaseous carbonyl form they may be separated through fractional distillation. This process is unlikely to yield a readily interpreted technosignature as high >99% purity metals and alloys will be produced on-demand (to minimise waste). The elimination of warehousing and waste is central to just-in-time manufacturing with high quality control procedures (lean manufacturing) [[131]]. Furthermore, carbonyls as gases will escape asteroid deposition. Carbonaceous material from C-type asteroids would offer a critical source of carbon for any number of applications. Silicone plastics are favoured over hydrocarbon plastics for their high durability and radiation and temperature tolerance. They may be manufactured from syngas ($CO+H_2$) and silicon feedstock using metal/mineral oxide catalysts through the Rochow process using recycled HCl [[132]] yielding no waste products. We now consider the processing of S-type asteroid resources, primarily for ceramic and other materials. The olivine fayalite ($Fe_2SiO_4$) reacts with water at 200-315°C yielding magnetite (magnetic ferrite), silica (source of high-grade fused silica glass) and hydrogen (reducing agent):

$3Fe_2SiO_4 + 2H_2O \rightarrow 2Fe_3O_4 + 3SiO_2 + 2H_2$

At 300°C and 500 bar, mixed olivines are serpentinised ($Mg_3Si_2O_5(OH)_4$) in water through magnetite catalysis while reducing $CO_2$ to $CH_4$:

$18Mg_2SiO_4 + 6Fe_2SiO_4 + 26H_2O + CO_2 \rightarrow 12Mg_3Si_2O_5(OH)_4 + 4Fe_3O_4 + CH_4$

Magnetite, silica and methane have useful applications as magnetic ferrite, a source of fused silica glass (amongst others such as the manufacture of piezoelectric quartz) and a source of syngas and as a reducing agent respectively. Since serpentinite is a natural product, none of these yield deposited waste products that can be identified as artificial. In the presence of quartz, forsterite ($Mg_2SiO_4$) reacts to form the orthopyroxene enstatite ($MgSiO_3$):

$Mg_2SiO_4 + SiO_2 \rightarrow 2MgSiO_3$

Pyroxene is dominant as enstatite in E-type asteroids. Enstatite ($MgSiO_3$) may be reacted with silicic acid ($H_4SiO_4$) at 200-500°C to form talc ($Mg_3Si_4O_{10}(OH)_2$):

---

[129] Ellery A (2024) "Vertical closure constraints on self-replicating machines" *Living Machines 2024*, Lecture Notes in Computer Science 14930, 269-283

[130] Sonter M (1997) "Technological and economic feasibility of mining the near-Earth asteroids" *Acta Astronautica* **41** (4-10), 637-647

[131] Liker J (2004) *Toyota Way: 14 Management Principles from the World's Greatest Manufacturer*, McGraw Hill Publishers

[132] Pakula D, Marciniec B, Przekop R (2023) "Direct synthesis of silicon compounds – from the beginning to the green chemistry revolution" *Applied Chemistry* **3** (1), 89-109

$3MgSiO_3 + SiO_2 + H_2O \rightarrow Mg_3Si_4O_{10}(OH)_2$

This requires a source of silica otherwise water will generate serpentinite. Talc has utility as a dry lubricant under high vacuum conditions where silicone oils are prone to evaporation. Under natural conditions, this requires hydrothermal fluids that may have occurred in protoplanetary nebula. This cannot be regarded as a technosignature. Enstatite can be reduced by $CH_4$ (from mixed olivine serpentinisation) at 1600°C to yield Si directly and syngas:

$MgSiO_3 + 2CH_4 \rightarrow Si + MgO + 2CO + 4H_2$

This recycles the syngas. MgO can be reduced directly to Mg using the molten salt electrolysis [[133]]. As a source of Si, molten salt electrolysis of silica will yield significantly higher purities of Si. Although less common than enstatite, the pyroxene augite ($Ca(Fe,Al)Si_2O_6$) may be artificially weathered with HCl to yield montmorillonite, $CaCl_2$ and $Fe(OH)_3$:

$3Ca(Fe,Al)Si_2O_6 + 10HCl + 5H_2O \rightarrow Ca(Al)_2(Si_4O_{10})(OH)_2.nH_2O + 2H_4SiO_4 + 2CaCl_2 + 3Fe(OH)_3$

Montmorillonite clay has widespread utility as the major component in bentonite drilling mud for use in excavating asteroids. $CaCl_2$ is the universal electrolyte in molten salt electrolysis in the industrial ecology (see later). It is plausible that asteroid mining will require significant amounts of drilling mud for asteroid drilling compared with molten salt electrolyte leading to an over-abundance of $CaCl_2$ as a potential technosignature. However, HCl may be recycled from $CaCl_2$-$H_2O$-rich brine at 380-500°C and 230-580 bar [[134]]: $CaCl_2 + 2H_2O \rightarrow Ca(OH)_2 + 2HCl$. At 512°C and 100 kPa, $Ca(OH)_2$ decomposes into CaO and water. In this case, there would be no excess $CaCl_2$ in evidence. Plagioclase (albite-anorthite series) feldspar is more common than alkali (orthoclase series) feldspar and both are associated with differentiated HED meteorites derived from 4-Vesta). Orthoclase may be treated with HCl acid:

$2KAlSi_3O_8 + 2HCl + 2H_2O \rightarrow Al_2Si_2O_5(OH)_4 + 2KCl + SiO_2 + H_2O$

This deposits the clay kaolinite which is used in the manufacture of porcelain for use as insulating junctions in electrical power distribution in the absence of polymer insulation (knob-and-tube technology). The discovery of porcelain products on asteroids would be of obvious artificial origin and invite an hypothesis of an alien "Beaker people".

We may search for the infrared spectra of extensive phyllosilicate deposits on S-type and C-type asteroids – such wastes would be expected from pre-processing of extraterrestrial minerals to extract specific metals or other materials using processes similar to chemical weathering on Earth [[135]]. C-type asteroids exhibit featureless low albedo <0.1 infrared spectra unlike phyllosilicate spectra which have an albedo of 0.4 expected from hydrated iron oxide minerals and an absorption feature at 0.7 μm associated with the OH feature at 3 μm. Other minerals include olivines, pyroxenes and feldspars from which phyllosilicates are derived. Evaporites such as $MgSO_4.7H_2O$ are common in CI and CM chondrites. Phyllosilicates can adsorb significant amounts of water between their octahedral and tetrahedral sheets that can be liberated by heating to 500-700°C. However, heating phyllosilicates to 1000-1100°C yielded similarities with C-type asteroid spectra by diminishing

[133] Ellery A, Mellor I, Wanjara P, Conti M (2022) "Metalysis FFC process as a strategic lunar in-situ resource utilisation technology" *New Space J* **10** (2), 224-238

[134] Bischoff J, Rosenbauer R, Fournier R (1990) "Generation of HCl in the system $CaCl_2$-$H_2O$: vapour-liquid relations from 380-500°C" *Geochimica et Cosmochimica Acta* **60** (1), 7-16

[135] Ellery A (2020) "Sustainable in-situ resource utilisation on the Moon" *Planetary & Space Science* **184** (4), 104870

absorption features at 1.4 μm and 2.3 μm and altering long-wavelength (1.8-2.5 μm) slopes [136]. Heating kaolinite and montmorillonite to >700°C yielded spectra resembling CI, CR, CK and CV chondrites while chlorite and serpentine yielded spectra resembling CM chondrites. The accompanying dehydration of 10-15% was associated with a 20-70% albedo drop consistent with the low albedo of C-type asteroids. This suggests that C-type asteroid surfaces are dominated by dehydrated phyllosilicates. Unless there is a means to differentiate artificially-generated phyllosilicates from naturally-generated phyllosilicates on C-type and S-type asteroids, this renders phyllosilicate detection on asteroids a challenging technosignature – see Fig 2 for pyroxene and orthoclase (lunar or asteroidal) processing yielding montmorillinite and kaolinite respectively on treatment with HCl acid. Silica, montmorillinite and kaolinite have direct engineering utility for a self-replicating machine. Hence, it is unlikely that evidence of chemical processing will be forthcoming on asteroids. In terms of manufacturing byproducts, additive manufacturing has virtually eliminated swarf waste [137]. The only residue that could be readily detected would be discarded machines and other manufacturing assets. However, given that we are concerned with self-replication capability which enforces material, energy and information efficiency [138], one would expect machinery to be repaired and exploited as self-repair is a subset of self-replication. Useful machinery will be incorporated into offspring probes or maintained to support local exploration of the target solar system. The latter of course may be extant and imply the existence of self-repairing sentinels within the asteroid population. The only other evidence may be in the scarring and fragmentation of asteroids – given that many asteroids are rubble piles, artificial scarring and fragmentation may be difficult to discern.

Lunar Industrial Ecology

Nevertheless, asteroids are not suitable for basing universal construction operations as the resources are distributed between different asteroids in different orbits and milli-gravity operations present significant challenges [139]. A larger gravitating body with some of the required bulk resources would be favourable to allow exploitation of gravity to control material handling – a large terrestrial planet moon. It seems not unreasonable to limit the ambient temperature of operations to that of habitability, i.e. the stellar habitable zone (SHZ) defined by liquid water. Firstly, stellar instellation is defined as:

$$S_* = \frac{R^2\, T^4}{M^{2/3} P^{4/3}}$$

where P=planetary period (y), $R = \frac{R_*}{R_\odot}$=stellar-to-solar radius ratio, $M = \frac{M_*}{M_\odot}$=stellar-to-solar mass ratio, $T = \frac{T_*}{T_\odot}$=stellar-to-solar effective temperature ratio. SHZ is defined around stars with stellar surface temperatures $T_{eff}$=4800-6300 K for mid-K to late-F spectral types with hotter stars having wider habitable zones than cooler stars. The size of the SHZ is given by the difference in semimajor axes of outer and inner limits of habitability:

---

136 Ostrowski D, Gietzen K, Lacy C, Sears D (2010) "Investigation of the presence and nature of phyllosilicates on the surfaces of C asteroids by an analysis of the continuum slopes in the near-infrared spectra" *Meteoritics & Planetary Science* **45** (4), 615-637

137 Jandyal A, Chaturvedi I, Wazir I, Raina A, Ul Haq I (2022) "3D printing – a review of processes, materials and applications in industry 4.0" *Sustainable Operations & Computers* **3**, 33-42

138 Ellery A (2024) "Vertical closure constraints on self-replicating machines" *Living Machines 2024*, Lecture Notes in Computer Science 14930, 269-283

139 Ellery A (2024) "Challenges of robotic milli-g operations for asteroid mining" *Proc Future Technologies Conference* **4** (ed. Arai K), Lecture Notes on Networks & Systems 1157, 45-64

$$\Delta a = a_{outer} - a_{inner} = RT^2\left(\frac{1}{\sqrt{S_{outer}}} - \frac{1}{\sqrt{S_{inner}}}\right)$$

From Kepler mission data of stellar dimming by planetary transits, it has been estimated that [140]: (i) 26±3% of sun-like GK stars have Earth-like planets (1-2$R_\oplus$) with orbital periods P=5-100 d; (ii) 1.6±0.4% of sun-like GK stars have hot Jupiter planets (8-16$R_\oplus$) with orbital periods P=5-100 d which is much less than (i); (iii) 5.7±2% of sun-like GK stars have Earth-like planets (1-2$R_\oplus$) with an Earth-like orbital period P=200-400 d; (iv) 22±8% of sun-like GK stars have Earth-like planets (1-2$R_\oplus$) within the SHZ of S=0.25-4.0$S_\odot$ which correlates to 0.5-2.0 AU (periods 130-1034 d) for the Sun where $S_\odot$=solar constant; (v) probability of an Earth-sized planet within 0.1-2.0 AU increases to ~50% for GK stars; (vi) 45% of M dwarfs have an Earth-sized planet (0.5-1.4$R_\oplus$) in the stellar habitable zone (0.25-1.5$S_\odot$). From the Kepler and Gaia databases, the number of terrestrial-sized planets $0.5R_\oplus<R<1.5R_\oplus$ in the SHZ around solar-type (mid-K to late F) stars with 4800 K<$T_{eff}$<6300 K is 0.37-0.60 (conservative SHZ limited by minimum and runaway greenhouse effects) or 0.58-0.88 (optimistic SHZ defined by Venus-Mars orbits) [141]. There is thus a habitable planet around a GK star (3900 K<$T_{eff}$<6000 K) within ~6 pc increasing to 4 such within ~10 pc of the sun with 95% confidence. Given the uncertainties inherent in limited sampling, it seems reasonable to assume that 50% GK stars possess a terrestrial-sized planet within their HZ yielding 287 million (conservative) habitable planets around G-stars within the Galaxy. The occurrence of terrestrial planets around any star is expected to be ~1 (we note that TRAPPIST-1 has seven rocky planets) and it is this that is relevant to the self-replicating probe in search of resources.

It has been estimated that, on average, 1 in 12 terrestrial planets possess a moon as large as or larger than our Moon generated by protoplanetary impacts [142]. Assuming a power or exponential law distribution of size imposed by natural physical processes, e.g. [143], smaller moons are expected to be much more common, perhaps to the extent that at least one terrestrial planet in the extrasolar system possesses a modest-sized moon. Indeed, the existence of a large moon has been highlighted as a potential anthropic condition for the existence of life on the primary companion [144]. Furthermore, such moons provide the stepping stones to solar system exploration evolving into interstellar migration. This is analogous to Pacific island-hopping by Polynesian ancestors from southeast Asia [145]. As the planets of our solar system become progressively more distant, so Pacific islands became progressively distant from southeast Asia. Given the Bowen mineral condensation sequence, such moons are expected have similar geologies to our own Moon. Our own Moon has been considered to host accumulations of alien artefacts, either delivered or detritus [146]. Alien

---

[140] Petigura E, Howard A, Marc G (2013) "Prevalence of Earth-sized planets orbiting sun-like stars" *Proc National Academy Sciences* **110** (48), 19273-19278

[141] Bryson S et al (2021) "Occurrence of rocky habitable-zone planets around solar-like stars from Kepler data" *Astronomical J* **161** (Jan), 36

[142] Elser S, Moore B, Studel J, Merishima R (2011) "How common are Earth-Moon planetary systems?" *Icarus* **214** (2), 357-365

[143] Golombek M and Rapp D (1997) "Size-frequency distributions of rocks on Mars and Earth analogue sites: implications for future landed missions," *J Geophysical Research* **102** (E2), 4117-4129

[144] Waltham D (2019) "Is Earth special?" *Earth-Science Reviews* **192** (May), 445-470

[145] Finney B (1988) "Solar system colonisation and interstellar migration" *Acta Astronautica* **18**, 225-230

[146] Arkhipov A (1998) "Earth-Moon system as a collection of alien artefacts" *J British Interplanetary Society* **51**, 181-184

artefacts on the Moon may be hidden in lava tubes with ~$10^8$ y longevity [[147]]. Our Moon is an obvious candidate for basing the universal constructor/self-replicator [[148]]. The Moon's partial gravity and its complementary material inventory compared with asteroids – aluminium, silicon, calcium and oxygen from anorthite in the highland regions [[149],[150]] and iron, titanium and oxygen from ilmenite in the mare regions [[151],[152]] – offer rich resources, particularly for bulk applications such as structure and other applications. Both sets of resources yield zero waste as our lunar industrial ecology recycles almost all waste [[153]]. Aluminium, extracted from abundant anorthite resources in the highlands, is a multifunctional material of enormous utility – lightweight structure, hardened tooling (silumin), electrically conducting wiring, thermal conducting straps, thermal radiators, highly reflective mirrors, thermite welding, major component of alnico permanent magnets, etc. Its oxide, alumina, has hardness and refractory properties second only to diamond. Although the mare basalt is not primordial, it is reasonable to expect significant late bombardment on extrasolar SHZ-inhabiting moons ~1.0-1.5 By after formation in conjunction with radiogenic heating yielding magma oceans in which incompatible elements are concentrated, e.g. Oceanus Procellarum. The existence of the magma ocean was crucial for the concentration of both thorium and uranium with KREEP minerals in Oceanus Procellarum [[154]]. To complement the Moon's resources, asteroids may be maneouvred and soft-landed onto the Moon using mass drivers [[155]]. We thus explore how our model of industrialisation of the Moon may proceed to understand what technosignatures of ETI visitation to search for.

---

[147] Arkhipov A (1995) "Search for alien artefacts on the Moon" *Progress in the Search for Extraterrestrial Life* (ed. Shostak S), ASP Conference series **74**, 259-264

[148] Ellery A (2018) "Engineering a lunar photolithoautotroph to thrive on the Moon – life or simulacrum?" *Int J Astrobiology* 17 (3), 258-280

[149] Thibodeau B, Walls X, Ellery A, Cousens B, Marczenko K (2024) "Extraction of silica and alumina from lunar highland simulant" *Proc ASCE Earth & Space Conf*, paper 6962

[150] Walls X, Ellery A, Marczenko K, Wanjara P (2024) "Aluminium metal extraction from lunar highland simulant using electrochemistry" *Proc ASCE Earth & Space Conf*, paper 7061

[151] Ellery A (2020) "Sustainable in-situ resource utilisation on the Moon" *Planetary & Space Science* **184** (4), 104870

[152] Ellery A, Mellor I, Wanjara P, Conti M (2022) "Metalysis FFC process as a strategic lunar in-situ resource utilisation technology" *New Space J* **10** (2), 224-238

[153] Ellery A (2021) "Are there biomimetic lessons from genetic regulatory networks for developing a lunar industrial ecology?" *Biomimetics J* **6** (3), 50

[154] Warren P, Wasson J (1979) "Origin of KREEP" *Reviews of Geophysics & Space Physics* **17** (1), 73-88

[155] Ellery A (2024) "Trials and tribulations of asteroid mining" *Proc ASCE Earth & Space Conf*, paper 8087

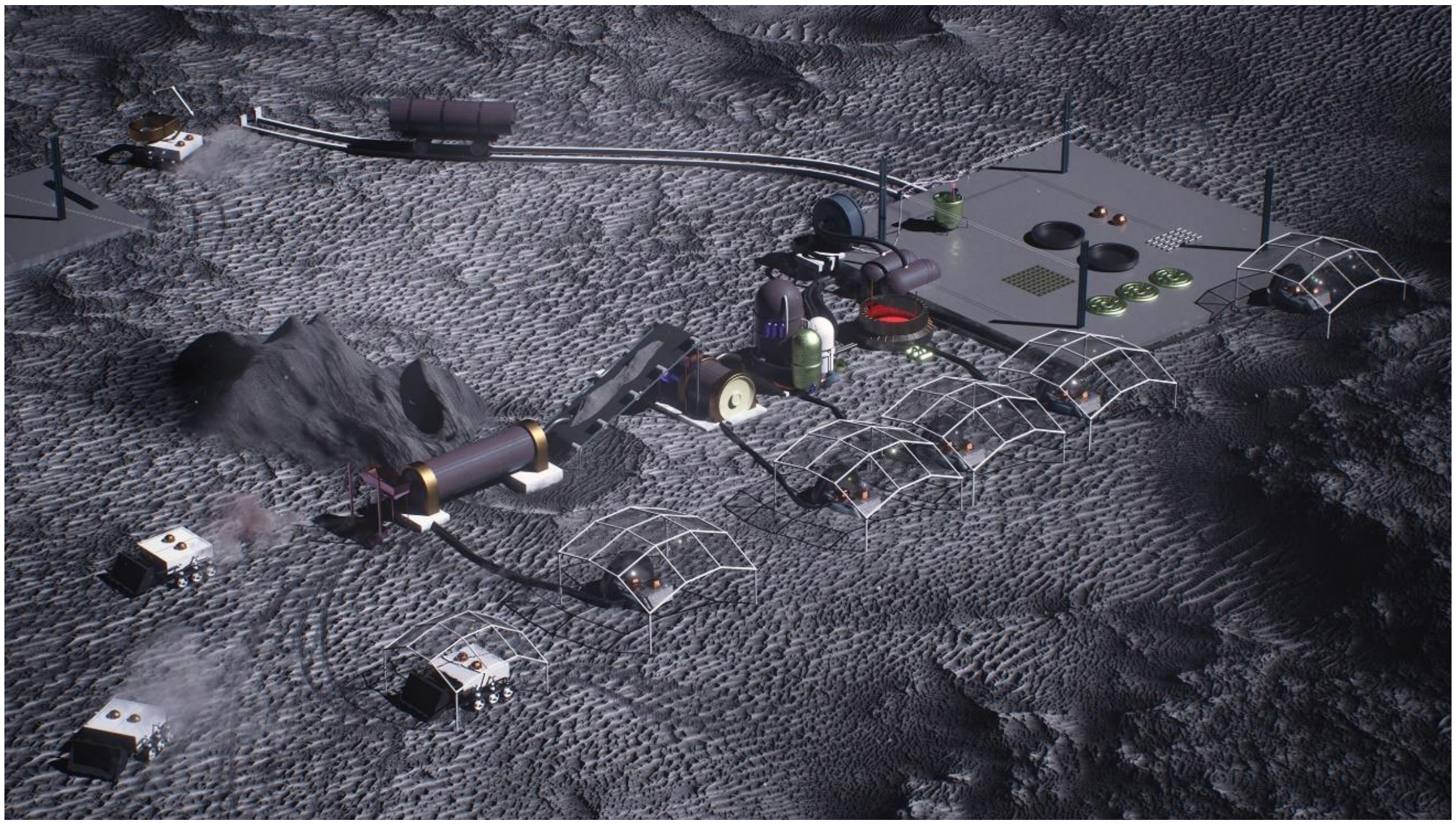

**Fig 2. Artist's rendition of a lunar industrial ecology implemented within a self-replicating system on the Moon**

For a lunar industrial ecology (Fig 3), we must adopt an Indigenous approach to leveraging the resources of the Moon (Fig 2) – we utilise everything and waste nothing. It is described in detail in [156] but the general philosophy is that the most appropriate technology will be as low-tech as possible to maximise incorporation of local resources into that technology with the exclusion of exotic, scarce and difficult-to-access materials. The simplest self-replicating machine can spawn more complex machines by virtue of its universal construction property [157]. This places a dividend on exploiting the simplest technology to enable physical closure of matter, energy and information through the replication cycle [158] – each component of the self-replicator embodies the history of its production processes including mining of its materials, its fabrication and assembly, the energy required to do so and the information required for its specification which imposes a physical cost of storage and processing [159]. An implication of this is we must define fundamental mechanisms to be re-used for multiple applications rather than developing a multiplicity of specific-to-task mechanisms (such exaptation is a ubiquitous constraint in biology [160]). This constrains design choices for the self-replicating machine's self-specification [161].

**Fig 3. Near-closed loop lunar industrial ecology (emboldened materials are pure metal oxides for direct reduction using the FFC molten salt electrolysis process)**

---

[156] Ellery A (2020) "Sustainable in-situ resource utilisation on the Moon" *Planetary & Space Science* **184** (4), 104870

[157] von Neumann J, Burks A (1966) "Theory of self-reproducing automata" *University of Illinois Press*, Urbana

[158] Ellery A (2024) "Vertical closure constraints on self-replicating machines" *Living Machines 2024*, Lecture Notes in Computer Science **14930**, 269-283

[159] Ellery A (2024) "Vertical closure constraints on self-replicating machines" *Living Machines 2024*, LNCS **14930**, 269-283

[160] Frenkel-Pinter M, Petrov A, Matange K, Travisano M, Glass J, Williams D (2022) "Adaptation and exaptation: from small molecules to feathers" *J Molecular Evolution* **90**, 166-175

[161] Ellery A (2022) "Lunar demandite – you gotta make this using nothing but that" *Proc ASCE Earth & Space Conf*, Colorado School of Mines, Denver, 743-758

Lunar Ilmenite

$Fe^0 + H_2O \rightarrow$ ferrofluidic sealing

$FeTiO_3 + H_2 \rightarrow \mathbf{TiO_2} + H_2O + Fe$

$2H_2O \rightarrow 2H_2 + O_2$

$2Fe + 1.5O_2 \rightarrow Fe_2O_3/Fe_2O_3.CoO$ - ferrite magnets

$3Fe_2O_3 + H_2 \leftrightarrow Fe_3O_4 + H_2O$) – formation of magnetite at 350-750°C/1-2 kbar

$4Fe_2O_3 + Fe \leftrightarrow 3Fe_3O_4$ )

Nickel-Iron Meteorites

W inclusions → Thermionic cathodic material

Mond process:

$W(CO)_6 \leftrightarrow 6CO + W$

$Fe(CO)_5 \leftrightarrow 5CO + Fe$ (175°C/100 bar) → Tool steel

$Ni(CO)_4 \leftrightarrow 4CO + Ni$ (55°C/1 bar) → Electrical steel

$Co_2(CO)_8 \leftrightarrow 8CO + 2Co$ (150°C/35 bar) → Permalloy

S catalyst

| Alloy | Ni | Co | Si | C | W |
|---|---|---|---|---|---|
| Tool steel | | | | 2% | 9-18% |
| Electrical steel | | | 3% | | |
| Permalloy | 80% | | | | |
| Kovar | 29% | 17% | 0.2% | 0.01% | |

$4FeS + 7O_2 \rightarrow \mathbf{2Fe_2O_3} + 4SO_2$

(Troilite) $SO_2 + H_2S \rightarrow 3S + H_2O$

$FeSe + Na_2CO_3 + 1.5O_2 \rightarrow \mathbf{FeO} + Na_2SeO_3 + CO_2$ at 650°C

$KNO_3$ catalyst $Na_2SeO_3 + H_2SO_4 \rightarrow Na_2O + H_2SO_4 + Se \rightarrow$ photosensitive Se

↑____________|

$Na_2O + H_2O \rightarrow 2NaOH$

$NaOH + HCl \rightarrow NaCl + H_2O$

Lunar Orthoclase

$3KAlSi_3O_8 + 2HCl + 12H_2O \rightarrow KAl_3Si_3O_{10}(OH)_2 + 6H_4SiO_4 + 2KCl$

orthoclase illite silicic acid (soluble silica)

$2KAl_3Si_3O_{10}(OH)_2 + 2HCl + 3H_2O \rightarrow 3Al_2Si_2O_5(OH)_4 + 2KCl$

kaolinite

$[2KAlSi_3O_8 + 2HCl + 2H_2O \rightarrow Al_2Si_2O_5(OH)_4 + 2KCl + \mathbf{SiO_2} + H_2O]$

$KCl + NaNO_3 \rightarrow NaCl + KNO_3$

$2KCl + Na_2SO_4 \rightarrow 2NaCl + K_2SO_4$

Lunar Olivine

$Fe_2SiO_4 + 4\ HCl + 4H_2O \rightarrow 2FeCl_2 + \mathbf{SiO_2} + 2H_2O$

fayalite

$Mg_2SiO_4 + 4HCl + 4H_2O \rightarrow 2MgCl_2 + 2H_4SiO_4$ → Sorel cement

forsterite $MgCl_2 + 2\ NaOH \rightarrow Mg(OH)_2 + 2NaCl$

$Mg(OH)_2 \rightarrow \mathbf{MgO} + H_2O$ → Sorel cement

600-800°C

Lunar Anorthite

$CaAl_2Si_2O_8 + 4C \rightarrow CO + \mathbf{CaO} + \mathbf{Al_2O_3} + 2Si$ at 1650°C → CaO cathode coatings

$CaO + H_2O \rightarrow Ca(OH)_2$

$Ca(OH)_2 + CO_2 \rightarrow CaCO_3 + H_2O$

$CaAl_2SiO_8 + 8HCl + 2H_2O \rightarrow CaCl_2 + 2AlCl_3.6H_2O + \mathbf{SiO_2}$ → fused silica glass + FFC electrolyte

$AlCl_3.6H_2O \rightarrow Al(OH)_3 + 3HCl + H_2O$ at 100°C

↑__________________________________________|

$2Al(OH)_3 \rightarrow \mathbf{Al_2O_3} + 3H_2O$ at 400°C → $2Al + Fe_2O_3 \rightarrow 2Fe + Al_2O_3$ (thermite)

AlNiCo hard magnets

Al solar sail

Lunar Pyroxene

$Ca(Fe,Al)Si_2O_6 + HCl + H_2O \rightarrow Ca_{0.33}(Al)_2(Si_4O_{10})(OH)_2.nH_2O + H_4SiO_4 + CaCl_2 + Fe(OH)_3$

Augite montmorillonite silicic acid iron hydroxide

Lunar Volatiles

$CO + 0.5\ O_2 \rightarrow CO_2$

$CO_2 + 4H_2 \rightarrow CH_4 + 2H_2O$ at 300°C (Sabatier reaction)

Ni catalyst

850°C 250°C

$CH_4 + H_2O \rightarrow CO + 3H_2 \rightarrow CH_3OH$ 350°C

Ni catalyst $Al_2O_3$ $CH_3OH + HCl \rightarrow CH_3Cl + H_2O$ 370°C $+nH_2O$

$Al_2O_3$ $CH_3Cl + Si \rightarrow (CH_3)_2SiCl_2 \rightarrow ((CH_3)_2SiO)_n + 2nHCl$ → silicone plastics/oils

↑________________________________________|

$N_2 + 3H_2 \rightarrow 2NH_3$ (Haber-Bosch process)

Fe on $CaO+SiO_2+Al_2O_3$

$4NH_3 + 5O_2 \rightarrow 4NO + 6H_2O$

WC on Ni

$3NO + H_2O \rightarrow 2HNO_3 + NO$ (Ostwald process)

↑________________|

$2SO_2 + O_2 \leftrightarrow 2SO_3$ (low temp)

$SO_3 + H_2O \rightarrow H_2SO_4$

<u>Salt Reagent</u>

$2NaCl + CaCO_3 \leftrightarrow Na_2CO_3 + CaCl_2$ (Solvay process) → FFC electrolyte

350°C/150 MPa

$Na_2CO_3 + SiO_2(i) \leftrightarrow Na_2SiO_3 + CO_2$ → piezoelectric quartz crystal growth (40-80 days)

1000-1100°C

$CaCO_3 \rightarrow$ **CaO** $+ CO_2$ (calcination)

$NaCl(s) + HNO_3(g) \rightarrow HCl(g) + NaNO_3(s)$

$2NaCl(s) + H_2SO_4(g) \rightarrow 2HCl(g) + Na_2SO_4(s)$

<u>Molten Salt (FFC) Process ($CaCl_2$ electrolyte)</u>

$MO_x + xCa \rightarrow M + xCaO \rightarrow M + xCa + ½xO_2$ where M=Fe, Ti, Al, Mg, Si, etc

$CO + 0.5\,O_2 \rightarrow CO_2$

$CO_2 + 4H_2 \rightarrow CH_4 + 2H_2O$ at 300°C (Sabatier reaction) → $CH_4 \rightarrow C + 2H_2$ at 1400°C → FFC anode regeneration

Ni catalyst

Consider the lunar industrial ecology that recycles all reagents as the basis of a self-replicating machine to be deployed on the Moon [162,163]. Closure constraints imposed by the self-replication cycle [164] require minimal use of exotic material and favouring of common rock-forming mineral resources from which multifunctional materials may be extracted [165]. The addition of asteroidal material permits full physical closure of the self-replication cycle. This industrial ecology can provide resources for the construction of lunar bases [166], lunar power architectures [167] and self-replication of modular factories [168]. The closure constraints favour simplicity over sophistication and, to that end, we propose that nuclear fission – leveraged from lunar resources - be deployed as the favoured in-situ power system to provide continuous power on-demand independent of lunar day-night cycles. We assume 2 $MW_e$ is required per fully-grown 20 tonne factory module (equivalent to a 100 m x 100 m solar array with 15% efficiency). The second task requires the construction of networks of surveying spacecraft for local extrasolar system exploration. Nuclear power provides the basis for nuclear-electric propulsion for local exploration of both inner and outer extrasolar systems with

---

162 Ellery A (2016) "Are self-replicating machines feasible?" *AIAA J Spacecraft & Rockets* **53** (2), 317-327

163 Ellery A (2024) "Self-replication technology for ubiquitous space exploration for all" *Int Astronautical Congress*, Milan, IAC-24-D4.1.9x80842

164 Ellery A (2024) "Vertical closure constraints on self-replicating machines" *Living Machines 2024*, LNCS 14930

165 Ellery A (2022) "Lunar demandite – you gotta make this using nothing but that" *Proc ASCE Earth & Space Conf*, Colorado School of Mines, Denver, 743-758

166 Ellery A (2021) "Leveraging in-situ resources for lunar base construction" *Canadian J Civil Engineering* **49** (5), 657-674

167 Ellery A (2021) "Generating and storing power on the Moon using in-situ resources" *Proc IMechE J Aerospace Engineering* **236** (6), 1045-1063

168 Ellery A (2024) "Self-replication technology for ubiquitous space exploration for all" *Int Astronautical Congress*, Milan, IAC-24-D4.1.9x80842 superior specific power (kg/kW) and efficient fuel consumption [169]. The adoption of nuclear fusion (requiring exotic and scarce fuels such as $^2H$ and/or $^3He$) is far more challenging than nuclear fission for the construction of local and space power plants. Nuclear fission could supply power to the self-replicating probe in-transit during long cruise phases when the interstellar propulsion system is disengaged. This suggests that any prior extraterrestrial visitation to the Moon entails that nuclear fission reactors may have been operated on the Moon.

Lunar Radioisotope Resources

High resolution mapping of lunar U (1764.5 keV) and Th (2614.5 keV) gamma rays have revealed an average ~0.3 ppm U (with large variations of up to 2 ppm especially at Procellarum KREEP basaltic terrane (PKT)) and ~1-2 ppm Th (with large variations of up to 7 ppm at PKT and average U/Th ratio of 0.27 similar to that of chondrites (0.28)) [170]. Similar measurements by Change-2's gamma ray spectrometer of Th abundance yields higher concentration in the eastern highlands than the western highlands especially within the Imbrium basin at PKT at >3 ppm and high concentrations up to 7.5 ppm Th due to the Fra Mauro Highlands [171]. Other elevated Th abundances at the South Pole Aitken (SPA) basin near the Ingenii basin are at 3.5 ppm compared with diminished abundance of ~1 ppm in feldspar-dominated highland terrain including the farside (except at the anomalous Compton-Belkovich region with 9 ppm) [172]. There may be much higher local concentrations of Th. U and Th are associated with KREEP minerals. Rare earth elements in lunar KREEP deposits have abundances of 180 ppm Nd and 65 ppm Dy while C-type asteroids exhibit a diminished 0.36-7.88 ppm Nd and 0.21-2.09 ppm Dy [173] thus favouring the Moon for such resources. Many of the KREEP minerals under Oceanus Procellarum and the Imbrium Basin on the nearside resemble their terrestrial counterparts but often with substitutions – apatite ($Ca_5(PO_4)(F,Cl)$), merrillite ($Ca_{17.3}Y_{0.4}(La\text{-}Lu)_{0.88}(Mg,Mn,Fe)_{2.4}(Na,K)_{0.07}(P,Al,Si)_{13.9}O_{56}$), monazite ($(REE,Th,Ca,Sr)(P,Si,S)O_4$), xenotime ($(REE,Zr)(P,Si)O_4$), yttrobetafite ($(Ca,Y,U,Th,Pb,REE)_2(Ti,Nb)_2O_7$ up to 9.45% Y) and tranquillityite ($Fe_8(Zr,Y)Ti_3Si_3O_{24}$ up to 4.6% Y and 0.25% Nd) [174]. Apatite, merrillite and monazite are the commonest rare earth sources and are found in association with each other in all KREEP deposits. Phosphates (apatite, whitlockite and merrillite) in stony meteorites as carriers of REE are depleted in REE compared with the Moon. KREEP minerals are enriched in U and Th – in yttrobetafite, U is in 38,200 ppm concentrations. Assuming a common terrestrial isotopic ratio $^{235}U/^{238}U$=0.72%, lunar U resources yield a $^{235}U$ peak concentration of 0.014 ppm so breeding is favoured over enrichment. Oceanus Procellarum exhibits around ~10 x $10^{-6}$ g/g of Th [175]. The abundance of thorium on the

[169] Niehoff J, Friedlander A (1974) "Comparison of advanced propulsion capabilities for future planetary missions" *J Spacecraft* **11** (8), 566-773

[170] Yamashita N, Hasebe N, Reedy R, Kobayashi S, Karouji Y, Hareyama M, Shibamura E, Kobayashi M-N, Okudaira O, d'Uston C, Grasnault O, Forni O, Kim K (2010) "Uranium on the Moon: global distribution and U/Th ratio" *Geophysical Research Letters* **37**, L10201

[171] Zhu M-H, Chang J, Ma T (2019) "Thorium distribution on the Moon: new insights from Chang'E-2 gamma ray spectrometer" *Research in Astronomy & Astrophysics* **19** (6), 76

[172] Zhang J, Liu J (2024) "Thorium anomaly on the lunar surface and its indicative meaning" *Acta Geochimica* **43**, 507-519

[173] Dallas J, Raval S, Saydam S, Dempster A (2021) "Investigating extraterrestrial bodies as a source of critical minerals for renewable energy technology" *Acta Astronautica* **186**, 74-86

[174] McLeod C, Krekeler M (2017) "Sources of extraterrestrial rare earth elements: to the Moon and beyond" *Resources* **6**, article 40

[175] Zou Y-L, Liu J-Z, Liu J-J, Xu T (2004) "Reflectance spectral characteristics of lunar surface materials" *Chinese J Astronomy & Astrophysics* **4** (1), 97-104

Moon offers the most obvious nuclear fuel by mining $ThO_2$ and transmuting it to fissile uranium [176]. Natural Th is almost 100% fertile $^{232}Th$ isotope with a half-life of 14 By. Th may be extracted chemically from monazite $(RE/U/Th)PO_4$ with 6-7% Th which is ground, for a series of treatments to precipitate out thoria $ThO_2$ [177]. This requires dissolution in NaOH at 400-500°C followed by treatment with hot concentrated HCl to dissolve rare earths. $Th(OH)_4$ is precipitated and calcined at 900-1000°C into $ThO_2$. Although this is more complex than the other lunar industrial ecology processes, NaOH and HCl reagents are already incorporated into the lunar industrial ecology (Fig 3). $ThO_2$ may be fabricated into fuel pellets or mixed with $UO_2$ or $PuO_2$ neutron sources - $^{232}Th$ must undergo slow neutron irradiation to decay into $^{233}U$: ${}^{232}_{90}Th(n,\gamma){}^{233}_{90}Th \xrightarrow{-\beta} {}^{233}_{91}Pa \xrightarrow{-\beta} {}^{233}_{92}U$ . This breeding process can be accomplished in a slow neutron thermal reactor with a neutron moderator. Given that U/Th extraction requires mining of KREEP minerals, some rare earth elements are suitable for reactor design applications such as Y and Yb which may be utilised in mineral form (e.g. xenotime) without requiring extraction from their native state.

Lunar Nuclear Fission Reactor

The Magnox reactor uses natural uranium fuel without enrichment within Mg-0.8Al-0.004Be (magnesium non-oxidising) alloy cladded fuel rods embedded in graphite bricks as the neutron moderator and hot $CO_2$ gas coolant at 4.3 MPa pressure imposing <450°C operating temperatures due to cladding tolerances. Carbon (for graphite moderator and $CO_2$ coolant) is available on the Moon but with low abundance ~100 ppm. Heating the organic fraction of carbonaceous chondrites at 700°C will release organic volatiles that may be oxidised in $O_2$ to $CO_2$ and water. Al may be sourced and extracted from lunar anorthite [178,179]. Mg has low neutron absorption cross section while Be reduces the tendency to oxidation – the 0.004% Be may be omitted because it serves to reduce oxidation sensitivity which may be compensated by adding small amounts of $CH_4$ into the coolant to prevent CO production – $CH_4$ may be generated from $CO_2$ at 400°C through the Sabatier reaction with an Ni-on-alumina catalyst as often proposed for Mars ($CO_2 + 4H_2 \rightarrow CH_4 + 2H_2O$) [180]. Impure lunar forsterite may be treated with HCl acid: $Mg_2SiO_4 + 4HCl + 4H_2O \rightarrow 2MgCl_2 + H_4SiO_4$. Silica may be precipitated from silicic acid. $MgCl_2$ may be treated with NaOH: $MgCl_2 + 2NaOH \rightarrow Mg(OH)_2 + 2NaCl$. $Mg(OH)_2$ heated to 600-800°C yields MgO: $Mg(OH)_2 \rightarrow MgO + H_2O$. MgO may be reduced directly into Mg metal through FFC molten salt electrolysis which is part of the lunar industrial ecology [181]. Thus, we may recover cladding material from lunar resources. Magnox control rods are 316 austenitic stainless steel (67% Fe, 18% Cr, 10% Ni, 2% Mo, 2% Mn and <1% S) which is an effective gamma ray shield. Fe may be sourced from the lunar mineral ilmenite ($FeTiO_3$) by hydrogen reduction and Ni from NiFe meteoritic material by the carbonyl process ($Ni + 4CO \leftrightarrow Ni(CO)_4$). For the critical properties of Cr, we require an extension of the lunar industrial ecology to incorporate Cr mining and

---

176 Schubert P, Marrs I, Daniel E, Conaway A, Bhaskaran A (2021) "Ultra-safe nuclear thermal rockets using lunar-derived fuel" *J Space Safety Engineering* **8**, 185-192

177 Schubert P, Marrs I, Daniel E, Conaway A, Bhaskaran A (2021) "Ultra-safe nuclear thermal rockets using lunar-derived fuel" *J Space Safety Engineering* **8**, 185-192

178 Thibodeau B, Walls X, Ellery A, Cousens B, Marczenko K (2024) "Extraction of silica and alumina from lunar highland simulant" *Proc ASCE Earth & Space Conf*, paper 6962

179 Walls X, Ellery A, Marczenko K, Wanjara P (2024) "Aluminium metal extraction from lunar highland simulant using electrochemistry" *Proc ASCE Earth & Space Conf*, paper 7061

180 Zubrin R, Muscatello A, Berggren D (2012) "Integrated Mars in-situ propellant production system" *ASCE J Aerospace Engineering* **26**, 43-56

181 Ellery A, Mellor I, Wanjara P, Conti M (2022) "Metalysis FFC process as a strategic lunar in-situ resource utilisation technology" *New Space J* **10** (2), 224-238 extraction from lunar chromite ($FeCr_2O_4$). This involves reacting with $Na_2CO_3$: $4FeCr_2O_4 + 8Na_2CO_3 + 7O_2 \rightarrow 8Na_2CrO_4 + 2Fe_2O_3 + 8CO_2$. Soluble $Na_2CrO_4$ is acifidied: $2Na_2CrO_4 + H_2SO_4 \rightarrow Na_2Cr_2O_7 + Na_2SO_4 + H_2O$. $Na_2Cr_2O_7$ is reduced with $SO_2$ to yield $Cr_2O_7$: $Na_2Cr_2O_7 + 3SO_2 + H_2O \rightarrow Cr_2O_3 + Na_2SO_4 + H_2SO_4$. $H_2SO_4$ is recycled and $Cr_2O_3$ may be reduced to Cr using through FFC molten salt electrolysis. S is readily sourced from asteroid-delivered troilite on the Moon ($4FeS + 7O_2 \rightarrow 2Fe_2O_3 + 4SO_2$ at 750-1100°C followed by $SO_2 + H_2S \rightarrow 3S + H_2O$). Mo and Mn sourcing and extraction require more challenging approaches because of their scarcity and complex extraction methods – Mo provides enhanced strength and resistance to corrosion at high temperatures while Mn increases hardness - both of which may be dispensed with through appropriate design accommodations. Control rod design may be improved by incorporating powdered Co (sourced from NiFe asteroids using the carbonyl process $2Co + 8CO \leftrightarrow Co_2(CO)_8$) within 316 steel tubes which act as "neutron windows". These may be mixed with ytterbium- and yttrium-containing rare earth minerals such as xenotime as neutron absorbers without extraction being necessary. Reactor vessel containment materials and pipework include corrosion-resistant 316 stainless steel among others rated to 20 bar [[182]]. Steels may also be mechanically alloyed oxide dispersion strengthening steels with 1% ultrafine oxide powder to reduce neutron embrittlement [[183]]. Silica and/or alumina (from lunar anorthite) oxide dispersed in stainless steel ensure the formation of protective oxide layers. Both steels are stable up to 650°C. Alternatives include Ti-5Al-2.4V with $Cr_2O_3$+$SiO_2$ or $TiO_2$ coating (up to 500°C) and SiC fibres reinforced SiC ceramic composite [[184]]. Silicon may be derived from anorthite as silica and reduced through FFC molten salt electrolysis to Si [[185]]. Reactor thermal radiator constructed from TiAl alloy (less V) is readily available from the lunar mineral ilmenite ($FeTiO_3$) following hydrogen reduction and molten salt electrolysis. Magnox was designed as a breeder reactor for generating $^{239}$Pu from fertile $^{238}$U so conversion to breading fertile $^{232}$Th into fissile $^{233}$U is straightforward. Rather than constructing reinforced concrete containment structures using lunar-sourced cements [[186]], lunar nuclear reactors could be buried with surface thermal radiators of steel with integrated coolant pipes to reject waste heat. Advanced gas cooled reactor improvements may be incorporated – graphite may be in pebble form rather than graphite blocks. A Brayton cycle with $CO_2$ working gas is more efficient than the Rankine cycle. The thermal efficiency of the Magnox reactor is 25-30% - the advanced gas cooled reactor with an efficiency of 40% operates at much higher temperature of 650°C requiring more exotic alloys which are much more challenging to source. Supercritical $CO_2$ is a fluid beyond its critical point that can be used as coolant up to 700°C yielding efficiencies ~50%. Supercritical $CO_2$'s corrosivity generates carburisation of steels requiring 316 austenitic stainless steels and nickel-based superalloys such as alloy 617 (including Co, Mo, W, Cr, Al) by forming a $Cr_2O_3$

---

[182] Ortner S (2023) "Review of structural material requirements and choices for nuclear power plant" *Frontiers in Nuclear Engineering* **2**, 1253974

[183] Ortner S (2023) "Review of structural material requirements and choices for nuclear power plant" *Frontiers in Nuclear Engineering* **2**, 1253974

[184] Victoria M, Baluc N, Spatig P (2020) "Structural materials for fusion reactors" *https://www-pub.iaea.org/mtcd/publications/pdf/csp_008c/pdf/ftp1_13.pdf*

[185] Ellery A, Mellor I, Wanjara P, Conti M (2022) "Metalysis FFC process as a strategic lunar in-situ resource utilisation technology" *New Space J* **10** (2), 224-238

[186] Mueller R, Howe S, Kochmann D, Ali H, Anderson C, Burgoyne H, Chambers W, Clinton R, De Kestellier X, Ebelt K, Gerner S, Hofmann D, Hogstrom K, Ives E, Jerves A, Keeman R, Keravala J, Khoshnevis B, Lim S, Metzger P, Meza L, Nakaruma T, Nelson A, Partridge H, Pettit D, Pyle R, Reiners E, Shapiro A, Singer R, Tan W-L, Vasquez N, Wilcox B, Zelhofer A (2016) "Automated additive construction (AAC) for Earth and space using in-situ resources" *Proc 15th Biennial ASCE Int Conf Engineering Science Construction & Operations in Challenging Environments (Earth & Space 2016),* Reston, VA

protective layer [187]. A small modular supercritical $CO_2$ gas fast reactor burning 12%-enriched $^{235}UO_2$ fuel at 650°C/20 MPa (near the critical point) offers 200 $MW_{th}$ with 50% thermal efficiency through the Brayton cycle [188]. Cladding was 316 stainless steel for corrosion-resistance up to 650°C. Their disadvantage is that supercritical $CO_2$ has poor neutron moderator performance requiring ZrH rods [189] and so sourcing and extraction of Zr on the Moon or asteroids would be challenging. Hence, there is an evolutionary path towards higher performance nuclear fission reactors on the Moon at the cost of broadening lunar industrial ecology diversity. The latter renders the former unattractive.

For lunar industrialisation, it requires just 6.5 years through 13 generations to construct a population of 1.5 M self-replicating factory modules on the Moon around the rims of Oceanus Procellarum, Mare Frigoris and Mare Imbrium within a 1 km wide membrane (assuming a six-month replication cycle for a 10 tonne seed factory module) [190]. Since each has a productive capacity of 60 tonnes per year (growth doubling and two offspring per generation), this yields a total productive throughput of 90 Mtonnes/y which exceeds US annual raw steel production (68 Mtonnes/y in 2024). At 2 $MW_e$ power requirement per factory module ($8 \times 10^{-5}$ g/s Th assuming 30% efficiency), this amounts to 3 $TW_e$ of nuclear power resources for the 1.5 M population of factories (120 g/s Th, i.e. almost 4000 tonnes Th/y compared with a total $^{232}$Th lunar crustal reserve of ~$10^{13}$ tonnes). This is almost exactly the Earth's annual electricity consumption (3.2 $TW_e$) as a K0.73 civilisation and the industrial throughput of 90 Mtonnes/y would take just under a month to build the dry mass of a Daedalus starship (much shorter than the time to fuel it [191]) for the offspring probe [192] but 160 years to build an O'Neill colony [193] after the offspring's launch (so this has no bearing on Galactic colonisation time). Fuelling of a Daedalus starship requires fleets of aerobots to mine Jupiter's atmosphere for its ~40 ppm $^3$He nuclear fusion fuel. Jupiter and its Sun-Jupiter libration points may be suitable staging posts for such aerobots. Jupiter's total inventory of $^3$He in the upper 10% of its atmosphere is $7 \times 10^{20}$ kg of which the Daedalus starship required $2 \times 10^7$ kg $^3$He which would take 20 years to mine with 1000 x 100 tonne aerobots (i.e. a productivity of 1000 kg/aerobot/y) [194]. If a deceleration stage were added for an interstellar encounter, this increases the fuel requirement by 38 times yielding $7.6 \times 10^8$ kg of $^3$He. This is a tiny fraction of Jupiter's total inventory and the detection of $^3$He depletion in Jupiter's atmosphere would require a $^3$He concentration measurement accuracy of ~$10^{-12}$. A fleet of 1.5 M x100 tonne aerobots could be manufactured within under 2 years by universal constructors to reduce the total $^3$He mining time to 6 months for a Daedalus-derived interstellar mission.

The discovery of the natural water-cooled uranium fission reactor at Oklo, Gabon that operated 1.8 By ago [195] was due to abnormal depletion in $^{235}U/^{238}U$ to 0.44-0.72% with a $^{232}$Th excess indicating

---

187 Lee J, Pint B (2021) "Corrosion in gas-cooled reactors" *Oak Ridge National Laboratory Report*, TLR-RES/DE/CIB-CMB-2021-04

188 Parma E, Wright S, Vernon M, Fleming D, Rochau G, Ahti S-A, Rashdan A, Tsvetkov P (2011) "Supercritical $CO_2$ direct cycle gas fast reactor (SC-GFR) concept" *Sandia National Laboratories report* SAND2011-2525

189 Wang L, Lu D, Yao L, Xiang H, Zhao C (2021) "Study on temperature feedback effect on supercritical $CO_2$-cooled reactor" *Frontiers in Energy Research* **9**, 2021.764906

190 Ellery A (2024) "Self-replication technology for ubiquitous space exploration for all" *Int Astronautical Congress*, Milan, IAC-24-D4.1.9x80842

191 Bond A, Martin A (1978) "Project Daedalus" *J British Interplanetary Society Supplement* **31**, S5-S7

192 Freitas R (1980) "Self-reproducing interstellar probe" *J British Interplanetary Society* **33**, 251-264

193 O'Neill G (1974) "Colonisation of space" *Physics Today* **27** (9), 32-40

194 Parkinson R (1978) "Project Daedalus: propellant acquisition techniques" *BIS Project Daedalus – Final Report*, S83-S89

195 Maurette M (1976) "Fossil nuclear reactors" *Annual Reviews Nuclear Science* **26**, 319-350

$^{235}U$ burning [196]. Soluble $UO_2$ in groundwater was deposited in sufficient concentrations at 3.54%, similar to that of light water reactor fuel enrichment, to become critical. Similarly, any anomalous $^{232}Th/^{144}Nd$ or $^{232}Th/^{137}Ba$ (assuming the $^{129}Xe$ gas branch dissipated due to lack of fluid inclusions) measurements on the Moon may be interpreted as evidence of previous artificial nuclear reactors burning $^{232}Th$. This will require detailed in-situ highly-sensitive isotopic measurements but ~TW energy generation for a self-replicating population of factories would yield a widespread isotopic signature. However, depleted isotope deposits will not necessarily coincide with their source location, e.g. the anomalous Compton-Belkovich region which lies on the other side of the Moon to PKT. A 50 km diameter Th-rich granite batholith has been discovered on the Moon buried below the Compton-Belkovich feature exhibiting anomalously high heat flux ~180 $mW/m^2$ (compared with average lunar highlands flux of 5-10 $mW/m^2$) generated by KREEP-based radiogenic remelting and crystal fractionation [197]. It has been suggested that fissile waste may be deposited into a star which could be detectable if deposited in sufficient quantities into a star with a shallow convective envelope to limit depth of mixing, i.e. A5-F2 only [198]. The wastes would be the result of slow neutron fission of $^{239}Pu$ and $^{233}U$ from breeder reactors burning $^{238}U$ and $^{232}Th$ respectively. In our case of fission of $^{233}U$ from $^{232}Th$ breeding, the waste would include $^{144}Nd$ and $^{137}Ba$ but the quantities would be small in relation to the $^{238}U$ and $^{232}Th$ reserves and the Sun's G2 convective envelope would impose deep mixing. This would render such solar dumping undetectable and eliminate $^{232}Th/^{144}Nd$ and $^{232}Th/^{137}Ba$ isotopic evidence of prior ET activity on the Moon. The considerable energy expenditure required to launch waste and its attendant spacecraft infrastructure into the Sun from the lunar surface without any advantage dictates against this scenario.

Lunar Artefacts

Anomaly (novelty) detection generally requires the adoption of a normal reference model and the use of a metric such as Mahalanobis distance or k-nearest neighbour for determining deviation from normal. Deep learning methods have demonstrated superior performance over more traditional methods including principal components analysis (PCA) and support vector machines (SVM) in dealing with the curse of dimensionality problem [199]. For learning systems, anomaly detection imposes a fundamental challenge, in that anomalies, by definition, are rare in datasets. This makes unsupervised learning methods more suitable for anomaly detection than supervised learning. There are several unsupervised learning approaches of which the most common are variational automatic encoders (VAE) and generative autoencoder networks (GAN) but there are a wide range of deep learning methods [200,201]. The autoencoder is a neural network comprising an encoding network and a decoding network through one or more narrow hidden layers [202]. The encoder compresses the input data and from that compressed form the decoder reconstructs the original data. Autoencoders are similar to PCA but are not restricted to linear dimensionality reduction. The VAE is a probabilistic extension of autoencoders that introduces regularisation using a prior distribution to prevent overfitting. VAE methods can result in high false positive rates. GAN comprise two convolutional neural networks [203] – a generator G to generate datasets that have the same distribution as the training data to fool the discriminator D into classifying as real data while D attempts to distinguish between training and synthetic data. They are trained with the objective function $\min_{G}\max_{D}\left(<\log\left(D(X)\right)>+<\log\left(1-D\left(G\left(\tilde{X}\right)\right)\right)>\right)$.

---

196 Kruglov A, Pchelkin V, Sviderskii M, Moshchanskaya N, Chernetsov O, Dymkov Y (176) "Natural nuclear reactor in Oklo Gabon" *Atomnaya Energiya* **41** (1), 3-9

197 Siegler M, Feng J, Lehman-Franco K, Andrews-Hanna J, Economos R, St Clair M, Million C, Head J, Glotch T, White M (2023) "Remote detection of a lunar granitic batholith at Compton-Belkovich" *Nature* **620** (Aug), 116-124

198 Whitmire D, Wright D (1980) "Nuclear waste spectrum as evidence of technological extraterrestrial civilisations" *Icarus* **42**, 149-156

199 Pang G, Shen C, Cao L, Van Den Hengel A (2020) "Deep learning for anomaly detection: a review" *ACM Computing Surveys* **54** (2), 1-38

200 Yang J, Xu R, Qi Z, Shi Y (2022) "Visual anomaly detection for images: a survey" *Procedia Computer Science* **199**, 471-478

201 Natha S, Leghari M, Rajput M, Zia S, Shabir J (2022) "Systematic review of anomaly detection using machine and deep learning techniques" *Quest Research J* **20** (1), 83-94

202 Chalapathy R, Chawla S (2019) "Deep learning for anomaly detection: a survey" *arXiv:1901.03407v2 [cs.LG]*

The narrow angle camera (NAC) on the Lunar Reconnaissance Orbiter has mapped the lunar surface to a resolution of 0.5 m in search of potential evidence of ETI in the form of durable artefacts or former mining activities [204]. Four deep learning methods – VAE, kernel density estimation, isolation forest and farpoint algorithm – were successfully applied to anomaly detection on LRO high resolution images of Ranger 6 crash site, Apollo 12 landing site, Apollo 13 Saturn IV-B crash site and Apollo 17 landing site as sample technosignatures [205]. The Apollo 15 region data set has been subjected to technosignature search for anomalies using unsupervised machine learning and successfully identified the Apollo 15 landing module as anomalous [206]. The unsupervised learning algorithm was a VAE comprising eight layers of alternating convolution and batch functions using KL divergence from a Gaussian prior. Deviation from fractal geometry over different scales of a visual feature may quantify its artificiality on the basis that natural features are fractal - the "face on Mars" feature was found to be non-fractal [207]. A variation on this is the likelihood ratio to differentiate between natural features (with fractal dimensions) and artificial features (with non-fractal dimensions) in images [208]. This suggested that there are regions of Mars that may be artificial - such results illustrate that automated techniques are not infallible and are crucially dependent on the quality, resolution and shadowing distortions in the images. For explainability to detect such errors, there are few options for implementing transparent anomaly detection [209]. Attention-based approaches capture gradients to compute attention weights which are used to explain novelties. Reasoning-based approaches utilise knowledge graphs for Bayesian inferencing but large language models (LMM) have demonstrated superior ability to reason about images. It has been suggested that our planetary missions have searched only a limited volume of the solar system at sufficient resolution to identify a ~10 m sized artefact [210]. The search volume is extremely large but the Earth and Moon have been mapped adequately to exclude surface artefacts. This does not rule out buried artefacts.

---

[203] Mohammadi B, Fathy M, Sabokrou M (2021) "Image/video deep anomaly detection: a survey" *arXiv:2103.01739v1 [cs.CV]*

[204] Davies P, Wagner R (2013) "Searching for alien artefacts on the Moon" *Acta Astronautica* **89**, 261-265

[205] Loveland R, Sime R (2024) "Anomaly detection methods for finding technosignatures" *Proc 13th Int Conf Pattern Recognition Applications & Methods*, 633-640

[206] Lesnikowski A, Bickel V, Angerhausen D (2020) "Unsupervised distribution learning for lunar surface anomaly detection" *arXiv:2001.04634v1 [astro-ph.IM] 14 Jan 2020*

[207] Carlotto M, Stein M (1990) "Method for searching for artificial objects on planetary surfaces" *J British Interplanetary Society* **43**, 209-216

[208] Carlotto M (2007) "Detecting patterns of a technological intelligence in remotely sensed imagery" *J British Interplanetary Society* **60**, 28-39

[209] Wang Y, Guo D, Li S, Camps O, Fu Y (2024) "Explainable anomaly detection in images and videos: a survey" *arXiv: 2302.06670v3 [cs.LG]*

[210] Haqq-Misra J, Kopparapu K (2012) "On the likelihood of non-terrestrial artefacts in the Solar System" *Acta Astronautica* **72**, 15-20

Economics as the rational distribution of scarce resources for the purpose of wealth creation by adding value is a universal mathematical concept [211] and exhibited in animal behaviour through the exchange of goods/services [212,213]. The former is based on rational and/or social choice theory [214,215] while the latter is biased with emotional behaviours such as gratitude which influence reciprocal exchanges [216]. In any social interaction, the tit-for-tat strategy is optimal and begins with cooperation [217]. Reciprocal exchanges include chimpanzees trading food sharing for sexual access. Fairness in trade is also critical. Capuchin monkeys react negatively to unequal reward distribution [218] and chimpanzees attempt to equalize such inequality. Animal conflicts also respect the asymmetry between prior territorial possession and territorial incursion by an interloper [219,220]. Hence, sensitivity to fair trade as an economic transaction is a biologically evolved capacity rather than a cultural one. Fair trade between a self-replicating probe and indigenous intelligent residents of the extrasolar system hosting resources is an obvious strategy for developing cooperative relations. The self-replicating probe thus categorically excludes the zoo hypothesis as a solution to the Fermi paradox (under certain conditions) [221,222]. Trade across interstellar distances is unlikely, but if implemented, transportation costs on interstellar goods should be based on transport time experienced by stationary observers rather than dilated time onboard the spacecraft [223]. For the self-replicating probe, transport costs do not apply - trade may be negotiated in-situ with resident intelligent species. To minimise costs to the residents of the extrasolar system, raw material extraction must be sustainable and not deprive the residents of critical resources – mining of exotic and scarce materials should not be undertaken, favouring the exploitation of common rock-forming minerals. Raw material in the extrasolar system for self-replication may be traded with: (a) information which may be scientific, technological or cultural; (b) physical product manufacture including O'Neill colonies; (c) the provision of space-based energy supplies [224]; (d) extrasolar system transportation facilities such as coilguns and/or nuclear-electric engines; (e) other specific technologies, the most valuable being universal construction and/or interstellar propulsion as the keys to evolution to KII/KIII civilisation-hood. If no resident intelligent species exists (which is the most likely scenario), an artefact which contains any of the above may be constructed as future payment to any future intelligent species that might evolve. This requires a long-term investment in future events that appear to be disfavoured by future time discounting in economic value. However,

---

211 Heilbroner R (1991) "Economics as universal science" *Social Research* **58** (2), 457-474
212 De Waal F (2021) "How animals do business" *Philosophical Trans Royal Society B* **376**, 20190663
213 Addessi E, Beran M, Bourgeois-Gironde S, Brosnan S, Leca J-B (2020) "Are the roots of human economic systems shared with non-human primates?" *Neuroscience & Biobehavioural Reviews* **109** (Feb), 1-15
214 Burns T, Roszkowska E (2016) "Rational choice theory: toward a psychological, social and material contextualisation of human choice behaviour" *Theoretical Economics Letters* **6** (2), 195-207
215 Myerson R (2013) "Fundamentals of social choice theory" *Quarterly J Political Science* **8** (3), 305-337
216 Axelrod R, Hamilton W (1981) "Evolution of cooperation" *Science* **211**, 1390-1396
217 Axelrod R (2006) *Evolution of Cooperation*, 2nd ed, Perseus Books, New York
218 Brosnan S, de Waal F (2003) "Monkeys reject unequal pay" *Nature* **425**, 297-299
219 Maynard Smith J, Price G (1973) "Logic of animal conflict" *Nature* **246**, 15-18
220 Parker G (1974) "Assessment strategy and the evolution of fighting behaviour" *J Theoretical Biology* **47**, 223-243
221 Ball J (1973) "Zoo hypothesis" *Icarus* **19** (3), 347-349
222 Crawford I, Schultze-Makuch D (2024) "Is the apparent absence of extraterrestrial technological civilisations down to the zoo hypothesis or nothing?" *Nature Astronomy* **8**, 44-49
223 Krugman P (2010) "Theory of interstellar trade" *Economic Inquiry* **48**, 1119-1123
224 Ellery A (2022) "Solar power satellites - rotary joints, magnetrons and all - from lunar resources?" *Proc ASCE Earth & Space Conf*, Colorado School of Mines, Denver, 773-788 there is little empirical evidence that human preferences are subject to a stable temporal discount rate [225]. This cannot be regarded as an effective counter to long-term investment strategies. Plundering resources of an extrasolar system contravenes the Outer Space Treaty (1967) regarding universal “celestial objects” – although the letter of the law favours humanity as it is human “common law”, the spirit of the law suggests equitability between parties [226]. Gift-exchange appears to be an ancient and universal human trait that imposes a social obligation of reciprocity [227]. Ostensibly, there is no expectation of equivalent return but there is overwhelming social pressure to reciprocate. The gift is a prototypical contract. If the perceived value of the artefact exceeds the cost of resources consumed (by the giver) to the receiver, this may be regarded as retroactive gift-giving from the giver that offers immediate excess beyond the payment price – the gift component - to the receiver on receipt. It is in this spirit that the consumption of resources by the giver should be traded with generosity encapsulated within the artefact. Such artefacts may include self-replicating probes or their products [228]. Universal construction technology would be such a gift in providing productive technology that offers: (a) quantity imparted by exponential growth in productive capacity; (b) quality imparted by universal construction ability to build any type of machine. Favoured locations for siting artefacts include Earth orbit, lunar orbit and orbits about Earth-Moon libration points [229]. Artificial robotic objects (lurkers) may be co-orbiting with near-Earth objects in highly elliptical horseshoe-shaped 1:1 resonant orbits and offer material resources, potential anchorage and camouflage [230]. However, their orbits are stable only for <My requiring periodic large $\Delta v$ manoeuvres comparable to that required for interplanetary transfer. These, and NEO and MBA populations, are excellent locations for a fleet of self-repairing deep-space sentinels to observe and explore unobtrusively over aeons. Powered by nuclear-electric propulsion to minimise fuel consumption, their $^{233}$U fuel has a half-life of 160,000 years so it is unlikely that periodic refuelling would be observable by humanity. Their electric propulsion fuel will require more frequent refuelling [231] but water and oxygen is plentifully and ubiquitously available in the solar system. Self-repair of sentinels, however, would be both material-specific and comparatively frequent, perhaps every few centuries, driven by the reliability requirements for interstellar transit times as a minimum duration of reliability. Given the earlier estimate of a single habitable planet around a GK star within ~6 pc (20 light years), this imposes a minimum of a 200-year reliability at an interstellar cruise velocity of 0.1c [232] which increases to ~300 years for a ~10 pc range (this assumes onboard maintenance and repair from onboard stock). Given the requirement for lunar resources, this requires maintenance of a lunar infrastructure such as mass drivers and their attendant power systems to launch from the lunar surface after self-repair. There is no evidence of any such infrastructure on the lunar surface. This may imply that there are no functional sentinels and any lunar infrastructure has been obliterated by micro-bombardment and burial through impact gardening over ~40-80 My on average up to ~300 My for some materials. It may be that maintenance

[225] Frederick S, Loewenstein G, O’Donoghue T (2002) “Time discounting and time preference: a critical review” *J Economic Literature* **40** (Jun), 351-401

[226] Lyall F, Larsen P (2020) *Space Law: A Treatise*, 2nd Ed, Routledge Publishing, Abingdon-on-Thames, UK

[227] Sherry J (1983) “Gift giving in anthropological perspective” *J Consumer Research* **10** (2), 157-168

[228] Freitas R (1983) “Search for extraterrestrial artefacts (SETA)” *J British Interplanetary Society* **36**, 501-506

[229] Freitas R (1983) “If they are here, where are they? Observational and search considerations” *Icarus* **55** (Aug), 337-343

[230] Benford J (2019) “Looking for lurkers: co-orbiters as SETI observables” *Astronomical J* **158** (Oct), article 150

[231] Tejeda J, Knoll A (2023) “Water vapour fuelled Hall effect thruster: characterisation and comparison with oxygen” *Acta Astronautica* **211**, 702-715

[232] Bryson S et al (2021) “Occurrence of rocky habitable-zone planets around solar-like stars from Kepler data” *Astronomical J* **161** (Jan), 36

of such infrastructure on the Moon for 300 years is more costly than rebuilding 1.5 M factory modules from the detritus of the old which can be achieved within only 6.5 years as required on-demand. More optimistically, artefacts as gifts may have been buried deliberately. We suggest that our Moon's subsurface should be the primary target for the search for buried artefacts and/or infrastructure by virtue of its stability over aeons and accessibility to a thresholded technological culture.

One potential location for buried artefacts is at locations where artefact technosignatures are masked from casual orbital survey but accessible to a species with demonstrable technological sophistication that suggests imminent expansion into interstellar space. The motive would be provision of the gift as a token of goodwill for future relations. The lunar industrial ecology requires "tunico" materials – tungsten, nickel and cobalt – which are found in abundance in M-type asteroidal material. Although high concentrations of Ni and Co in metal particles of lunar regolith are common [233], they are too diffusely distributed to be a useful resource due to impact gardening over the aeons. The most promising locations are where concentrated asteroidal resources have been deposited into the Moon's surface by impacts [234,235]. A large 100 km diameter impactor at an oblique angle will survive as an iron-rich impact deposit [236] and such large impacts generate transient magnetic fields which may be frozen into mineral signatures [237]. Indeed, oblique impact by a 200 km diameter iron-rich asteroid at 45$^o$ led to the distribution of magnetic anomalies ~10 nT of impactor material at the northern rim of the SPA basin [238]. It is conceivable that M-type asteroid delivery may have been deliberately soft-landed onto the lunar surface [239] by self-replicating probes that were subsequently obliterated through micro-bombardment. This would yield a surface signature of enriched deposits but would lack the impact cratering of natural hard impacts. Near infrared/gamma ray spectroscopy has yielded ambiguous results during a magnetic survey of the upper few cm of regolith at the South Pole Aitken (SPA) basin [240]. These indirect estimates of $\delta FeO=FeO_{GRS}-FeO_{NIR}$ as a measure of metallic iron did not detect metallic iron within <10 cm depth of the surface up to the resolution limits of the instruments. However, natural delivery of M-type asteroid material through shallow-angle impacts would be expected to yield a more diffuse surface signature or burial of such a signature. Since the formation of SPA ~4.3 By ago, we would expect subsequent impact gardening to have overlain any metallic iron deposited. Burying an artefact in association with these resources would be an ideal location to mask it with natural M-type signatures. Access to such subsurface M-type material will require extensive mining capabilities whether it be open-pit, underground tunnelling or in-situ drilling. This pre-supposes an extensive lunar technological infrastructure to find such an artefact – a threshold of technological capability that can maximise the utility of the artefact.

---

[233] Wittman A, Korotev R (2013) "Iron-nickel(-cobalt) metal in lunar rocks revisited" *44th Lunar & Planetary Science Conf*, abstract 3035

[234] Bland P, Cintala M, Horz F, Cressey G (2001) "Survivability of meteorite projectiles – results from impact experiments" *32nd Lunar & Planetary Science Conf*. abstract no 1764

[235] Bland P, Artemieva N, Collins G, Bottke W, Bussey D, Joy K (2008) "Asteroids on the Moon: projectile survival during low velocity impacts" *39th Lunar & Planetary Science Conf*, abstract no 2045

[236] Wakita S, Johnson B, Garrick-Bethell I, Kelley M, Maxwell R, Davison T (2021) "Impactor material records the ancient lunar magnetic field in antipodal anomalies" *Nature Communications*, 12.6543

[237] Crawford D (2020) "Simulations of magnetic fields produced by asteroid impact: possible implications for planetary palaeomagnetism" *Int J Impact Engineering* **137**, 103464

[238] Wieczorek M, Weiss B, Stewart S (2012) "Impactor origin for lunar magnetic anomalies" *Science* **335**, 1212-1215

[239] Ellery A (2024) "Trials and tribulations of asteroid mining" *Proc ASCE Earth & Space Conf*, paper 8087

[240] Cahill J, Hagerty J, Lawrence D, Klima R, Blewett D (2014) "Surveying the South Pole Aitken basin magnetic anomaly for remnant impactor metallic iron" *Icarus* **243**, 27-30

In offering self-replication technology, the artefact offers the lowest cost means to implement lunar infrastructure [ [241] ]. Mastery of self-replication technology on the Moon implies the cusp of interstellar capability – interstellar flight requires the productive capacity afforded by self-replication [[242]]. We suggest that during our efforts at lunar industrialisation, a byproduct may be the discovery of evidence of ETI and resolution of the Fermi paradox. We refer to this as the Vygotsky hypothesis to illustrate the intimate dependence of the physical technology of tools on the development of language and reasoning [[243]].

## Conclusions

It has been argued that, given the limits of our exploration of our volumetrically vast solar system, it is unsurprising that we have not yet discovered evidence of alien robotic probes given their small sizes ~1-10 m [[244]]. We have presented potential locations for targeted searches and admitted that technosignatures could be larger but hidden, though there may be widespread isotopic evidence of artificial nuclear power generation. Our search for such technosignatures will be side-effects of our own efforts to industrialise lunar and asteroid resources. The anomalous Chicxulub crater had been discovered by petroleum geologists decades before its significance was revealed by Alvarez et al [[245]]. For the Moon and asteroids, we are forearmed. The quicker we get to it, the quicker we may discover the answer to one of the most important scientific and philosophical questions of our time. In closing, we have been somewhat anthropocentric in focussing on the inner solar system in our search for solar system technosignatures. Our experience informs our science but our experience may not be that of ETI – the universe is stranger than we can imagine according to Jack Haldane.

---

[241] Ellery A (2017) "Space exploration through self-replication technology compensates for discounting in NPV cost-benefit analysis – a business case?" *New Space J* **5** (3), 141-154

[242] Ellery A (2022) "Self-replicating probes are imminent – implications for SETI" *Int J Astrobiology* **21** (4), 212-242

[243] Calvin W (1993) "Unitary hypothesis: a common neural circuitry for novel manipulations, language, plan-ahead and throwing?" in *Tools, Language and Cognition in Human Evolution* (ed. Gibson K, Ingold T), Cambridge University Press, 230-250

[244] Haqq-Misra J, Kopparapu K (2012) "On the likelihood of non-terrestrial artefacts in the solar system" *Acta Astronautica* **72**, 15-20

[245] Alvarez L, Alvarez W, Asaro F, Michel H (1980) "Extraterrestrial cause for the Cretaceous-Tertiary extinction" *Science* **208**, 1095-1108